\documentclass[preprint,12pt]{elsarticle}

\biboptions{sort&compress}

\usepackage{booktabs} 
\usepackage{mathident}
\usepackage{amsmath}
\usepackage{graphicx}
\usepackage{listings}
\usepackage{subfig}
\usepackage{xcolor}
\usepackage[ruled,linesnumbered]{algorithm2e}
\usepackage{multirow}
\usepackage{url}
\usepackage{pythonhighlight}

\newcommand{\kw}[1]{\mathtt{#1}}

\newcommand{\jss}[1]{\textcolor{black}{{#1}}}

\journal{Systems and Software}

\begin{document}

\begin{frontmatter}

\title{Flexible Control Flow Graph Alignment for Delivering Data-Driven Feedback to Novice Programming Learners}


\author[1]{Md Towhidul Absar Chowdhury}
\ead{mac9908@rit.edu}
\author[1]{Maheen Riaz Contractor}
\ead{mc1927@rit.edu}
\author[1]{Carlos R.~Rivero}
\ead{crr@cs.rit.edu}

\affiliation[1]{organization={Rochester Institute of Technology},
            addressline={One Lomb Memorial Dr.}, 
            city={Rochester},
            postcode={14623}, 
            state={NY},
            country={USA}}

\begin{abstract}
Supporting learners in introductory programming assignments at scale is a necessity. This support includes automated feedback on what learners did incorrectly. Existing approaches cast the problem as automatically repairing learners' incorrect programs extrapolating the data from an existing correct program from other learners. However, such approaches are limited because they only compare programs with similar control flow and order of statements. A potentially valuable set of repair feedback from flexible comparisons is thus missing. In this paper, we present several modifications to CLARA, a data-driven automated repair approach that is open source, to deal with real-world introductory programs. We extend CLARA’s abstract syntax tree processor to handle common introductory programming constructs. Additionally, we propose a flexible alignment algorithm over control flow graphs where we enrich nodes with semantic annotations extracted from programs using operations and calls. Using this alignment, we modify an incorrect program's control flow graph to match correct programs to apply CLARA’s original repair process. We evaluate our approach against a baseline on the twenty most popular programming problems in Codeforces. Our results indicate that flexible alignment has a significantly higher percentage of successful repairs at 46\% compared to 5\% for baseline CLARA. Our implementation is available at \url{https://github.com/towhidabsar/clara}.
\end{abstract}




\end{frontmatter}


\section{Introduction}

The worldwide interest in computer science has originated an unprecedented growth in the number of novice programming learners in both traditional and online settings~\cite{DBLP:conf/sigcse/GarciaCDDR16, Rodriguez2012, DBLP:conf/pldi/WangSS18, ZwebenB16}. In the latter case, the number of novices taking programming Massive Open Online Courses and/or practicing using programming online judges has scaled to millions~\cite{DBLP:conf/aied/HuangPNG13, DBLP:journals/apin/ToledoM17}. One of the main challenges in the aforementioned context is supporting novice programming learners at scale~\cite{DBLP:conf/pldi/SinghGS13}, which typically consists of delivering feedback explaining what and why they did incorrectly in their programs~\cite{KirschnerSC06}. Note that, different than traditional settings, online programming settings often have a large proportion of novice learners with a variety of backgrounds, who usually tend to need a more direct level of feedback and assistance~\cite{DBLP:conf/cscw/CoetzeeFHH14}. A common practice to address such a challenge is to rely on functional tests; however, feedback generated based solely on test cases does not sufficiently support novice learners~\cite{DBLP:conf/pldi/GulwaniRZ18, DBLP:conf/pldi/SinghGS13}.

Current approaches cast the problem of delivering feedback to novices at scale as automatically repairing their incorrect programs~\cite{DBLP:conf/pldi/GulwaniRZ18, DBLP:conf/icml/PiechHNPSG15, DBLP:conf/oopsla/PuNSB16, DBLP:conf/icse/RolimSDPGGSH17, DBLP:conf/pldi/SinghGS13, DBLP:conf/pldi/WangSS18}. Note that, similar to existing approaches, we consider a program to be correct if it passes a number of predefined test cases~\cite{DBLP:conf/pldi/GulwaniRZ18, DBLP:conf/pldi/WangSS18}; otherwise, it is incorrect. Once a repair is found, it can be used to determine pieces of feedback to deliver to learners~\cite{DBLP:conf/pldi/SinghGS13}. Non-data-driven approaches aim to find repairs by mutating incorrect programs until they are correct, i.e., they pass all test cases~\cite{DBLP:journals/csur/Monperrus18}. Data-driven approaches exploit the fact that repairs can be found in existing correct programs and extrapolated to a given incorrect program~\cite{DBLP:conf/pldi/WangSS18}. This paper focuses on the latter since, in a given programming assignment, there is usually a variety of correct programs provided by other learners that can be exploited to repair incorrect programs~\cite{DBLP:conf/pldi/GulwaniRZ18, DBLP:conf/oopsla/PuNSB16, DBLP:conf/icse/RolimSDPGGSH17, DBLP:conf/pldi/WangSS18}.

The ``search, align and repair''~\cite{DBLP:conf/pldi/WangSS18} framework consists of the following steps: 1)~Given an incorrect program $p_i$, search for a correct program $p_c$ that may be useful to repair $p_i$; 2)~Align $p_i$ with respect to $p_c$ to identify discrepancies and potential modifications in order to repair $p_i$; and 3)~Apply those modifications to $p_i$ until the resulting program $p_i'$ passes all test cases. Current approaches instantiating the ``search, align and repair'' framework use rigid comparisons to align incorrect and correct programs, i.e., they require the programs to have the same or very similar control flows (conditions and loops), and they are affected by the order of program statements~\cite{DBLP:conf/pldi/GulwaniRZ18, DBLP:conf/icse/RolimSDPGGSH17, DBLP:conf/pldi/WangSS18, DBLP:conf/oopsla/PuNSB16}. As a result, such approaches may miss a potentially valuable set of correct programs that can repair incorrect programs using flexible program comparisons.

In this paper, we focus on CLARA~\cite{DBLP:conf/pldi/GulwaniRZ18}, a ``search, align and repair'' approach that is open source. We first adapt the original implementation of CLARA to support introductory programming assignments. This adaption involves non-trivial modifications to the parser and interpreter to support various constructs, such as print to and read from the console, import statements, built-in Python functions, and more. After these modifications, we also need to adapt the alignment and repair processes. Based on these foundations, we propose a flexible alignment algorithm that relies on control flow graphs. It exploits the semantic information (operations and calls) to annotate the graphs, and their topology information (edges, i.e., True and False transitions). In order to evaluate the proposed algorithm, we create a dataset of incorrect and correct programs for the twenty most popular programming problems in the Codeforces online platform. Then, using the dataset, we execute CLARA's baseline repair process and our flexible alignment repair to compare both the quantitative and qualitative performance of the proposed technique. Furthermore, we include another ``search, align and repair'' approach, Sarfgen~\cite{DBLP:conf/pldi/WangSS18}, by utilizing a similar process as our flexible alignment, but enforcing a high similarity between compared programs. This simulates the rigidity in program comparisons applied by Sarfgen. Note that Sarfgen is not publicly available; therefore, we needed to simulate it.

Two short versions of this paper have been published elsewhere~\cite{DBLP:conf/its/MarinCR21, DBLP:conf/its/ContractorR22}. In this paper, we describe in detail all the modifications that we made to CLARA, and how the parser and interpreter were updated. We also present our flexible alignment algorithm as well as the changes made to the programs at hand after an alignment is computed. These changes are necessary in order to apply CLARA's repair process. Finally, we have significantly expanded our experiments to show the performance of our modifications over twenty real-world introductory programming assignments. We have made the implementation of this version of CLARA and our experimental dataset publicly available.\footnote{\url{https://github.com/towhidabsar/clara} The fundamental contributions of this paper are as follows:~1)~Many of the modifications we made to CLARA's parser and interpreter for Python programs also apply to other programming languages like C and Java;~2)~Both our flexible alignment and model recreation algorithms can be used by any ``search, align and repair'' approach based on control flow graphs and program expressions;~3)~Our dataset is one of the very few publicly-available datasets in the context of real-world introductory programming assignments;~4)~Our threshold-based solution to simulate other alignment approaches is useful when existing approaches are not publicly available.}

The paper is organized as follows: Section~\ref{sec:overview} summarizes previous approaches and ours; Section~\ref{sec:intro-clara} introduces CLARA's parser, interpreter, aligner and repairer; Sections~\ref{sec:modifications} and~\ref{sec:mod-repair} describe our modifications to the parser and interpreter, and aligner and repairer of CLARA's original implementation, respectively; Section~\ref{sec:flex-align} presents our flexible alignment approach and the necessary modifications to CLARA's repairer; Section~\ref{sec:evaluation} discusses our experimental results; Section~\ref{sec:related} presents the related work; and Section~\ref{sec:conclusions} presents our conclusions and future work. 
\section{Overview}
\label{sec:overview}

We consider CLARA~\cite{DBLP:conf/pldi/GulwaniRZ18}, Refazer~\cite{DBLP:conf/icse/RolimSDPGGSH17}, Sarfgen~\cite{DBLP:conf/pldi/WangSS18}, and sk\_p~\cite{DBLP:conf/oopsla/PuNSB16} the state of the art in searching, aligning and repairing programs. CLARA and Sarfgen compare variable traces between an incorrect and a correct programs that share the same control statements like \texttt{if} or \texttt{while}. Refazer uses pairs of incorrect/correct program samples to learn transformation rules, which aid a program synthesizer to transform incorrect into correct programs. Finally, sk\_p uses partial fragments of contiguous statements to train a neural network to predict possible repairs.

\newsavebox{\correctListing}
\begin{lrbox}{\correctListing}
\begin{minipage}[b]{0.4\textwidth}
\lstset{numbers=left, stepnumber=1,escapechar=;,}
\begin{python}
a = list(map(int, 
    input().split(' ')))
x, m, s = 0, 101, 0
while x < len(a):;\label{line:while-correct-killer};
    s += a[x]
    if m > a[x]:;\label{line:if-correct-killer};
        m = a[x]
    x+=1;\label{line:end-if-correct-killer};
print(s, ',', m)
\end{python}
\end{minipage}
\end{lrbox}

\newsavebox{\incorrectListing}
\begin{lrbox}{\incorrectListing}
\begin{minipage}[b]{0.4\textwidth}
\lstset{numbers=left, stepnumber=1,escapechar=;,}
\begin{python}
a = list(map(int, 
    input().split(' ')))
i, m, s = 0, 101, 9
while i < len(a):
    if m < a[i]:
        m = a[i]
    s += a[i]
    i+=1
    if m == 0:;\label{line:extra-if-init-incorrect-killer};
        i-=1;\label{line:extra-if-end-incorrect-killer};
print(s, ',', m)
\end{python}
\end{minipage}
\end{lrbox}

\newsavebox{\fragOneListing}
\begin{lrbox}{\fragOneListing}
\begin{minipage}[b]{0.3\textwidth}
\begin{python}
while x < len(a):
    s += a[x]
    if m > a[x]:
\end{python}
\end{minipage}
\end{lrbox}

\newsavebox{\fragTwoListing}
\begin{lrbox}{\fragTwoListing}
\begin{minipage}[b]{0.3\textwidth}
\begin{python}
while x < len(a):
    if m > a[x]:
        m = a[x]
\end{python}
\end{minipage}
\end{lrbox}

\begin{figure}
\centering
\subfloat[Correct program]{\usebox{\correctListing}\label{fig:correct-killer}}\qquad\qquad
\subfloat[Incorrect program]{\usebox{\incorrectListing}\label{fig:incorrect-killer}}\\
\subfloat[Excerpt of (simplified) abstract syntax tree edits]{\includegraphics[scale=0.85]{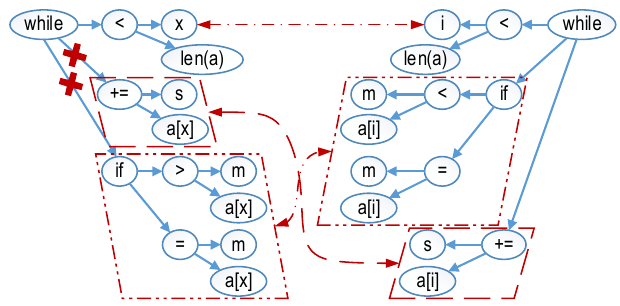}\label{fig:ast-diff-killer}}\\
\subfloat[Lines~\ref{line:while-correct-killer}--~\ref{line:if-correct-killer} in Figure~\ref{fig:correct-killer}]{\quad\quad\usebox{\fragOneListing}\quad\quad\label{fig:frag-one-killer}}~
\subfloat[Fragment needed to fix Figure~\ref{fig:incorrect-killer}]{\quad\quad\usebox{\fragTwoListing}\quad\quad\label{fig:frag-two-killer}}
\caption{Correct and incorrect programs, edits of abstract syntax trees derived from the programs and code fragments}
\label{fig:killer-example}
\end{figure}

In the alignment step, these approaches compare an incorrect program with respect to a correct program based on rigid schemes, which limits their repair potential. To illustrate our claim, we use the Python programs presented in Figure~\ref{fig:killer-example}, which aim to compute the minimum value in an array and the sum of all its elements, and print both minimum and sum values to console. Note that the values of the input array are assumed to be always less or equal than 100. In Sarfgen, an incorrect program will be only repaired if its control statements match with the control statements of an existing correct program. This is a hard constraint since:~a)~It requires a correct program with the same control statements to exist, and~b)~Such a correct program may not ``naturally'' exist. For instance, the control statements of the correct program in Figure~\ref{fig:correct-killer} do not match with the incorrect program in Figure~\ref{fig:incorrect-killer}; in order to match, the correct program should ``artificially'' contain an \texttt{if} statement before or after line~\ref{line:end-if-correct-killer}, and such a statement should not modify the final output of the program. CLARA relaxes these constraints such that, outside loop statements, both programs can have different control statements, but they need to have the same inside loops. This relaxation still forces a correct program with the same loop signature to exist.

Refazer exploits the tree edit distance between two programs to find discrepancies between them; however, the tree edit distance between two equivalent abstract syntax trees with different order of statements implies multiple edits. For example, Figure~\ref{fig:ast-diff-killer} shows an excerpt of the edits to transform the abstract syntax tree of the correct into the incorrect program in our example, which implies removing and adding full subtrees; however, only two edits would be necessary, i.e., changing ``$<$'' by ``$>$'' and removing the subtree formed by lines~\ref{line:extra-if-init-incorrect-killer}--\ref{line:extra-if-end-incorrect-killer} in Figure~\ref{fig:incorrect-killer}. In sk\_p, different order of statements result in different partial fragments, so additional correct programs will be required to train the program repairer. For instance, Figure~\ref{fig:frag-one-killer} shows a fragment extracted from the correct program; however, the incorrect program will only be fixed by a fragment like the one in Figure~\ref{fig:frag-two-killer}.

We propose an alignment step based on flexible alignment of control flow graphs. The first step consists of transforming programs into control flow graphs that encode the True and False transitions of the program at hand. For instance, the \texttt{while} loop in Figure~\ref{fig:correct-killer} (line~\ref{line:while-correct-killer}) is encoded by three nodes in the graph: the guard, the body and the end of the loop. There are transitions (edges) between these nodes. For example, a True transition between the guard and the body encodes that the guard is fulfilled, so the body is executed. These nodes are further annotated with semantic labels. For example, the guard of the loop contains the following labels: $cond$, indicating it is a Boolean condition, $Lt$, because there is a less than operator, and $len$, which corresponds to the \texttt{len} call. We apply flexible graph alignment over two (correct and incorrect) control flow graphs $G_C$ and $G_I$. Assume a given permutation of nodes $\phi$ such that every node $u_i \in G_C$ corresponds to a node $v_j \in G_I$, i.e., $\phi(u_i)=v_j$. We compute the similarity of $\phi$ as the similarity between the labels of $u_i$ and $v_j$ (semantic similarity), and the transitions (edges) outgoing from $u_i$ and $v_j$ (topology similarity). We select the permutation with lowest similarity as the best alignment.

\section{Introduction to CLARA~\cite{DBLP:conf/pldi/GulwaniRZ18}}
\label{sec:intro-clara}

\begin{figure}
\centering
\includegraphics[scale=.55]{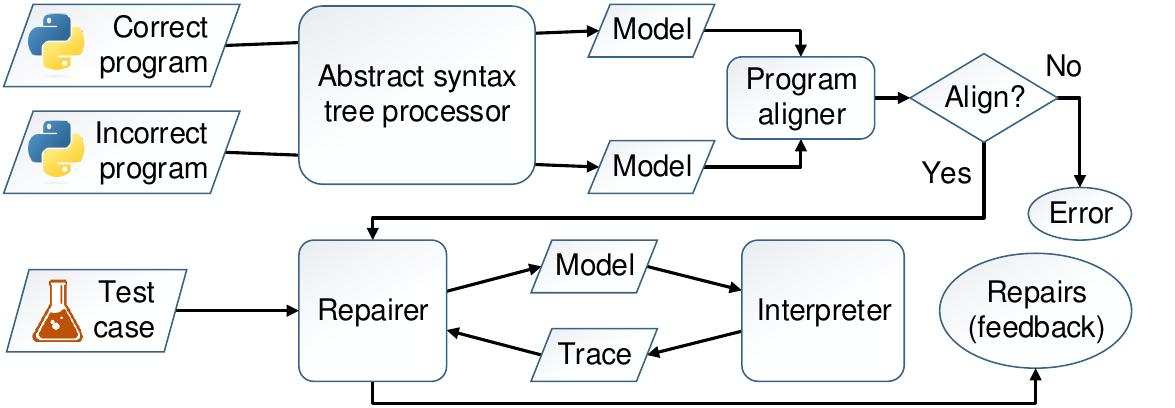}
\caption{CLARA's workflow: each program is translated into a model. Both models are aligned. If they match, the repairer and the interpreter exchange model and trace information using a test case until the incorrect program's model passes such test case.}
\label{fig:workflow}
\end{figure}

The automated CLustering And program RepAir tool, CLARA~\cite{DBLP:conf/pldi/GulwaniRZ18}, was first introduced in 2016. CLARA helps provide feedback to students in introductory programming assignments. Even though CLARA supports C++, Java and Python, we focus on the latter in this paper. Figure \ref{fig:workflow} presents CLARA's workflow. The abstract syntax tree processor receives each correct and incorrect programs as input, parses them, and creates two models, one for each program. These models are aligned: if a match is found, the repairer uses it as well as a test case to find potential errors and fix them. Error detection is achieved by comparing variable traces between the correct and incorrect programs.


\subsection{Models, processing and interpreting}

Before performing any repairs, CLARA creates models for every input program. CLARA exploits Python's $\kw{ast}$ module, which is a standard library that helps generate abstract syntax trees from Python source code and manipulate them. An abstract syntax tree is a graphical representation of a piece of source code containing nodes, where every node represents a language construct or operation like $\kw{If}$, $\kw{Return}$ and $\kw{Import}$~\cite{cruz_2021}. CLARA traverses the returned abstract syntax tree, node by node, and creates a model. For every part of the abstract syntax tree, such as $\kw{FunctionDef}$ (function definition), $\kw{Expr}$ (expression) or $\kw{Call}$ (function call), there is a different processing function that has a different representation in the model. 

\paragraph{Functions and locations} The entire model consists of one program containing multiple functions. Each function is partitioned based on its control flow information. Control flow information captures the order in which statements are evaluated. An example of a control flow statement is an $\kw{If}$ statement, as it adds a new possible path for the program to take. Locations, a construct in CLARA's model, represent control flow information. Each function thus contains multiple locations. Locations contain expressions and are created based on branching control flow statements like $\kw{If}$, $\kw{While}$ or $\kw{For}$ statements. Other statements like function calls or sequencing of statements are not involved in location creation. As a result, if a function contains no branching control flow statements, it will only contain a single location. Locations also contain information about which location to go to next, called transitions. There are two types of transitions, True and False. A True transition contains the following location to go to if the conditional expression inside the location evaluates to true. A False transition contains the location to visit next if the expression inside the location evaluates to false.

However, it is possible to have expressions inside a location that do not evaluate to a Boolean value, in which case, they will always go to a specific location. For example, transitioning back to the program body after executing a $\kw{Then}$ or an $\kw{Else}$ branch within an $\kw{If}$ statement. In this case, these are always True transitions. To maintain consistency, these locations also have True and False transitions, but the False transition always points to $\kw{None}$, and the True transition always points to the next location.

\paragraph{Expressions} CLARA contains three types of expressions as follows, where the names between parentheses refer to CLARA's naming convention: variables ($\kw{Var}$), operators ($\kw{Op}$), and constants ($\kw{Const}$). A $\kw{Const}$ can be a string, byte, number, or a name constant. A $\kw{Var}$ represents variables and, therefore, only comprises strings. An $\kw{Op}$ is the most complex type of expression as it encompasses all computations involving any type of operation, such as creating a list, set or tuple, and computations involving comparisons, if conditions, or binary operations. Every $\kw{Op}$ comprises two components, the name of the operation and the arguments it has to operate on. Depending on the type of operation, $\kw{Op}$ can contain a different number of arguments. For example, the $\kw{GetElement}$ operator entails getting an element from a list, dictionary or set, and contains two arguments: the object it needs to get the element from and the element index. $\kw{SetInit}$, which creates a set, has multiple arguments as each argument is an element inside the set. 

All types of control flow statements are also operators. However, while processing those statements, CLARA makes changes to the model. 
As mentioned earlier, processing control flow statements results in the addition of new locations to the model. The number of locations is different for each control flow statement. If the program contains an $\kw{If}$ statement, three or four locations will be added: one for the condition of the statement, another for the expressions inside the $\kw{Then}$ branch, one for the expressions after the statement, and, finally, another for the expressions inside the $\kw{Else}$ branch. The latter location is optional as we do not always have an $\kw{Else}$ branch accompanying the $\kw{Then}$ branch. However, CLARA recursively applies the following optimization to improve program comparison: an $\kw{If}$ statement that has no loops within is translated into a ternary operator. As a result, this statement is embedded in its parent and does not add any new nodes to the control flow graph. On the other hand, loops always result in the addition of three locations, corresponding to the guard, body, and the statements after the loop.

CLARA restricts its models by requiring a variable to appear only once on the left side of an expression per location. 
This restriction entails inspecting all the declarations of a particular variable and nesting the declarations in the last use of the variable. Hence, during the repair step, where variable traces are compared between models, there is only one value per location, making the comparison deterministic.

\paragraph{Example of a model}

\newsavebox{\pythonSampleSourceListing}
\begin{lrbox}{\pythonSampleSourceListing}
\begin{minipage}[b]{0.20\textwidth}
\lstset{numbers=left,stepnumber=1,escapechar=;,}
\begin{python}
a = [5, 6]
b, c = a ;\label{program-line:sample-source-b1};
b += 1 ;\label{program-line:sample-source-b2};
b += c ;\label{program-line:sample-source-b3};

for i in a:
    c += i
\end{python}
\end{minipage}
\end{lrbox}

\begin{figure}
\centering
\usebox{\pythonSampleSourceListing}
\caption{Sample Python source code}
\label{fig:PythonSourceCodeForModel}
\end{figure}
  
\begin{figure}
\centering
\small
\begin{lstlisting}[numbers=none,keywordstyle=\color{black},stringstyle=\color{black}, commentstyle=\color{black},xleftmargin=0.1\textwidth, columns=fullflexible]
Loc 1 (around the beginning of function main)
-----------------------------------
  a := ListInit(5, 6)
  c := GetElement(a', 1)
  b := AssAdd(AssAdd(GetElement(a', 0), 1), c')
  iter#0 := a'
  ind#0 := 0
-----------------------------------
  True -> 2, False -> None

Loc 2 (the condition of the 'for' loop at line 6)
-----------------------------------
  $cond := Lt(ind#0, len(iter#0))
-----------------------------------
  True -> 4, False -> 3

Loc 3 (*after* the 'for' loop starting at line 6)
-----------------------------------
-----------------------------------
  True -> None, False -> None

Loc 4 (inside the body of the 'for' loop beginning at line 7)
-----------------------------------
  i := GetElement(iter#0, ind#0)
  ind#0 := Add(ind#0, 1)
  c := AssAdd(c, i')
-----------------------------------
  True -> 2, False -> None
\end{lstlisting}
\caption{Model generated by CLARA for the program in Figure~\ref{fig:PythonSourceCodeForModel}}
\label{fig:PythonModelForSourceCode}
\end{figure}

Figures~\ref{fig:PythonSourceCodeForModel} and~\ref{fig:PythonModelForSourceCode} present an example of Python source code and its corresponding model. This model is the pretty-printed version created by CLARA to help improve readability.
Since the source code contains a $\kw{For}$ loop, the model contains four locations. The True and False transitions are shown at the bottom of each location and indicate how to traverse the model. For example, if $\kw{\$cond}$ in location 2 evaluates to true, we transition to location 4. If it evaluates to false, we transition to location 3. $\kw{ind\#0}$ corresponds to the index of the loop, $\kw{i}$, and $\kw{iter\#0}$ corresponds to the value we are iterating over, $\kw{a}$. CLARA internally creates both variables for every loop. As it can be seen in the source code, $\kw{b}$ appears on the left side of an expression three times (lines~\ref{program-line:sample-source-b1},~\ref{program-line:sample-source-b2} and~\ref{program-line:sample-source-b3}), but in the model, it only appears once in location 1. All three uses are nested inside one expression: $\kw{b := AssAdd(AssAdd(GetElement(a', 0), 1), c')}$. The other expressions correspond to the other operations before the loop. Whenever CLARA uses a variable defined earlier in an expression, it converts it to a different variable rather than using the original variable. For example, $\kw{a}$ becomes $\kw{a'}$. CLARA uses this notation (prime) to determine whether a variable is being defined or used.

Since CLARA relies on program execution and variable traces, it exploits a Python interpreter that helps execute models. The interpreter visits each expression in the model and recursively executes it, since an expression can contain other nested expressions. For example, the expression $\kw{b := AssAdd(AssAdd(GetElement(a', 0), 1), c')}$ evaluates from the innermost expression, $\kw{GetElement(a', 0)}$, until the most external expression. The interpreter contains functions for every type of operator, in this case, $\kw{AssAdd}$ and $\kw{GetElement}$, annotated with the word $\kw{execute\_}$ as a prefix for each function. Therefore, while evaluating the previous expression, CLARA first executes $\kw{GetElement(a', 0)}$, which translates to getting the element of $\kw{a}$ at index 0, invoking the function $\kw{execute\_GetElement()}$. It then evaluates $\kw{AssAdd(..., 1)}$ that increments the result by 1 using $\kw{execute\_AssAdd()}$, and so on.

\subsection{Single function alignment and program repair}

To find an alignment between two programs, CLARA creates models for each of them. If the control flow of both models is the same, for every variable in one model, it finds a matching with a corresponding variable in the other model based on variable tracing. The process consists of comparing values of variables at each location of the program trace. If the matching variables hold the same values at every point in the trace, the two programs are aligned. Single program repair is performed between two programs, where one is a correct program and the other is incorrect. The repair process starts by creating models for each program and aligning them as described above. Additionally, CLARA requires the programs to have at least one function and the same number of functions overall in the programs being compared. These functions must also have the same names and cannot be nested.  These features determine the structure of the programs. If there is a difference in the structure of the programs, CLARA throws a structure mismatch error and does not run.

\begin{algorithm}
\SetKwInOut{Input}{in}
\SetKwInOut{Inout}{in/out}
\SetKwInOut{Output}{out}
\Input{$G_C = (U, E)$ control flow graph (correct program); $u \in U$; $G_I = (V, F)$ control flow graph (incorrect program); $v \in V$}
\Inout{$\phi: U \rightarrow V$ program alignment}
\Output{Whether there is an alignment}
// If $u$ and/or $v$ are in $\phi$, they must be mapped to each other. \\
\If{$u \in dom~\phi \lor v \in ran~\phi$} { \label{alg:cap-if-u-v-in-phi}
    \Return $\phi(u) = v$ \label{alg:cap-return-u-v-in-phi}
}
// Add $u$ and $v$ to $\phi$. \\
$\phi(u) \gets v$ \\ \label{alg:cap-add-u-v-to-phi}
// $u'$ and $u''$ ($v'$ and $v''$) are the neighbors of $u$ ($v$). \\
Let $\{u \xrightarrow{True} u', u \xrightarrow{False} u''\} \subseteq E$, $\{v \xrightarrow{True} v', v \xrightarrow{False} v''\} \subseteq F$ \\
// Align $u$ and $v$ neighbors. \\
\Return $Align(G_C, u', G_I, v', \phi) \land Align(G_C, u'', G_I, v'', \phi)$ \label{alg:cap-recursive}
\caption{Align}
\label{alg:cap}
\end{algorithm}

Algorithm~\ref{alg:cap} corresponds to CLARA's program alignment based on the control flow graphs (locations) of the programs at hand. The algorithm receives two control flow graphs, $G_C = (U, E)$ and $G_I = (V, F)$, where $U$ and $V$ are sets of locations, and $E$ and $F$ are sets of transitions, and two locations $u \in U$ and $v \in V$, respectively. In the initial call, $u$ and $v$ are the entry points of both programs. The algorithm receives $\phi$, which determines the current mapping of $U$ locations into $V$ locations, i.e., the alignment between the graphs. It outputs whether or not there is an alignment between the graphs. If any of the locations is present in $\phi$, it returns whether they are both mapped (lines~\ref{alg:cap-if-u-v-in-phi}--\ref{alg:cap-return-u-v-in-phi}). Note that, if they are mapped, there is a match; otherwise, either $u$ or $v$ are mapped to a different location and, therefore, there is a mismatch. If $u$ and $v$ are not in $\phi$, they are added to $\phi$ (line~\ref{alg:cap-add-u-v-to-phi}), and their neighbors are recursively inspected (line~\ref{alg:cap-recursive}).



CLARA exploits $\phi$ during the repair process. For every variable in one location of the correct program, CLARA aims to match such variable in the corresponding location of the incorrect program. Due to CLARA's modeling, a variable can only have one expression per location, ensuring it can only hold one value. This makes the comparison with other variables possible. During the repair process, a mapping of a variable only deals with its specific expression and the expression it is mapped to. While mapping a variable from the correct program, every variable in the incorrect program is compared, whether or not it has already been mapped. Two variables are declared a match if their corresponding expressions evaluate to the same values using the same inputs, where the inputs denote the variables both the expressions depend on. This match is evaluated based on cost. Since both expressions depend on different variables, the cost of a match denotes the number of changes/steps it takes to transform the incorrect program's expression into the correct program's expression, where a variable from the incorrect program replaces each variable in the correct program's expression.

In the repair process, CLARA assumes that the number of variables in the correct program is the minimum number of variables needed, so the number of variables in the incorrect program needs to match the number of variables in the correct program precisely. Therefore, if the incorrect program contains extra variables, CLARA suggests deleting the variables it cannot match. If the incorrect program contains fewer variables than the correct program, CLARA suggests creating new variables. During cost calculation, all the variables in the correct program's expression are substituted by variables from the incorrect program as follows: Assume $\kw{s~=~a~+~2}$ is a statement in the correct program. CLARA replaces $\kw{a}$ by all other variables present in the corresponding location in the incorrect program and evaluates the associated cost. If the incorrect program contains three variables $\kw{x}$, $\kw{y}$ and $\kw{z}$ in the corresponding location, $\kw{a}$ in $\kw{s~=~a~+~2}$ is replaced by $\kw{x}$, $\kw{y}$ and $\kw{z}$, respectively. Since it is possible that the correct program has extra variables, CLARA also calculates the cost of the substitution using a fresh variable that does not exist in the incorrect program. This value is initialized to the value of $\kw{a}$ with a cost of 1. CLARA finally saves all the costs for every location and provides the cost array to a linear programming solver, which minimizes the overall cost and suggests matches for all the variables involved. Based on these matches and costs, the final set of repairs are suggested, which can be of three types: variable additions, variable deletions, or variable changes.

\newsavebox{\ModelsForCost}
\begin{lrbox}{\ModelsForCost}
\begin{minipage}[b]{0.90\textwidth}
\begin{minipage}{0.45\linewidth}
\centering
\textbf{Correct Model:}
\small
\begin{lstlisting}[columns=fullflexible, numbers=none,xleftmargin=0.3\textwidth, stringstyle=\color{black},]
a := 1
b := 2
c := Add(a', 1)
\end{lstlisting}
\end{minipage}\qquad
\begin{minipage}{0.45\textwidth}
\centering
\textbf{Incorrect Model:}
\small
\begin{lstlisting}[numbers=none, xleftmargin=0.3\textwidth,stringstyle=\color{black},columns=fullflexible]
x := 1
y := 2
z := Add(y', 1)
\end{lstlisting}
\end{minipage}
\end{minipage}
\end{lrbox}

\newsavebox{\CostTableC}
\begin{lrbox}{\CostTableC}
\small
\begin{tabular}[t]{|c|c|c|c|c|}
\hline
\# & Variable (C)
 & Variable (IC) & Dependency & Cost \\
\hline
1 & c & * & (a, *) & 4\\
\hline
2 & c & * & (a, x) & 4\\
\hline
3 & c & * & (a, y) & 4\\ 
\hline
4 & c & * & (a, z) & 4\\
\hline
5 & c & x & (a, *) & 2\\
\hline
6 & c & x & (a, y) & 2\\
\hline
7 & c & x & (a, z) & 2\\
\hline
8 & c & y & (a, *) & 3\\
\hline
9 & c & y & (a, x) & 3\\
\hline
10 & c & y & None & 0\\
\hline
11 & c & y & (a, z) & 3\\
\hline
12 & c & z & (a, *) & 1\\
\hline
13 & c & z & (a, x) & 1\\
\hline
14 & c & z & (a, y) & 0\\
\hline
\end{tabular}
\end{lrbox}

\newsavebox{\CostTableA}
\begin{lrbox}{\CostTableA}
\small
\begin{tabular}[t]{|c|c|c|c|c|}
\hline
\# & Variable (C)
 & Variable (IC) & Dependency & Cost \\
\hline
1 & a & * & None & 2\\
\hline
2 & a & x & None & 0\\
\hline
3 & a & y & None & 1\\ 
\hline
4 & a & z & None & 2\\
\hline
\end{tabular}
\end{lrbox}

\newsavebox{\CostTableB}
\begin{lrbox}{\CostTableB}
\small
\begin{tabular}[t]{|c|c|c|c|c|}
\hline
\# & Variable (C)
 & Variable (IC) & Dependency & Cost \\
\hline
1 & b & * & None & 2\\
\hline
2 & b & x & None & 1\\
\hline
3 & b & y & None & 0\\ 
\hline
4 & b & z & (a, y) & 0\\
\hline
5 & b & z & None & 3\\
\hline
\end{tabular}
\end{lrbox}

\begin{figure}
\centering
\usebox{\ModelsForCost}
\caption{Sample correct and incorrect program models}
\label{fig:ModelsCorrectAndIncorrect}
\end{figure}

\begin{table}
\centering
\caption{Repair cost table for variable $\kw{a}$ in Figure \ref{fig:ModelsCorrectAndIncorrect}}
\usebox{\CostTableA}
\label{fig:RepairCostTableForA}
\end{table}

We use the models shown in Figure~\ref{fig:ModelsCorrectAndIncorrect} to illustrate how the variable matching process works. Both models contain a single location that are trivially mapped to each other.  Table~\ref{fig:RepairCostTableForA} presents the variable matching of $\kw{a}$ in the correct program. The table contains five columns as follows: The first column represents the variable comparison number. The second column is the variable in the correct program we aim to match. The third column represents the possible variable match in the incorrect program. The fourth column contains the possible substitution for the dependent variables as a tuple $\kw{(i,j)}$, where $\kw{i}$ is the dependent variable from the correct program, and $\kw{j}$ is its possible substitution in the incorrect program. The fifth column is the cost for each variable match. In the first row, $\kw{*}$ entails that a fresh variable is used; the cost is 2 because we need to create a new variable and assign its value to 1 since $\kw{a}$'s value is one. The cost of the second row is 0 because no changes are needed. Note that there are no variable dependencies in this example.

\begin{table}
\centering
\caption{Repair cost table for variable $\kw{b}$ in Figure \ref{fig:ModelsCorrectAndIncorrect}}
\usebox{\CostTableB}
\label{fig:RepairCostTableForB}
\end{table}

\begin{table}
\centering
\caption{Repair cost table for variable $\kw{c}$ in Figure \ref{fig:ModelsCorrectAndIncorrect}}
\usebox{\CostTableC}
\label{fig:RepairCostTableForC}
\end{table}

Table~\ref{fig:RepairCostTableForB} shows the cost computation for variable $\kw{b}$. Note that the fourth row consists of replacing $\kw{b}$ by $\kw{z}$, which depends on $\kw{y}$; therefore, additional combinations of variables in the correct program are used, e.g., $\kw{y}$ and $\kw{a}$ are matched. Finally, Table~\ref{fig:RepairCostTableForC} presents the cost computation for variable $\kw{c}$ in which fresh variables are also used in the dependencies. The optimal solution is a matching that minimizes the overall cost for all variables. All of these possible matches for every variable are the input provided to the linear programming solver. The best variable matching is as follows: $\{\phi(a)=x, \phi(b)=y, \phi(c)=z\}$, which corresponds to rows 2, 3 and 14 in Tables~\ref{fig:RepairCostTableForA},~\ref{fig:RepairCostTableForB} and~\ref{fig:RepairCostTableForC}, respectively. The overall cost is 1 and the suggested repair is as follows: Replace $\kw{z~:=~Add(y', 1)}$ by $\kw{z~:=~Add(x', 1)}$ with cost = 1.0.
\section{Parser and interpreter modifications}
\label{sec:modifications}

We analyzed introductory programming assignments to identify language constructs commonly used like print statements, input functions and import statements. Some of these statements were not supported by CLARA's original implementation. In this section, we report the language constructs we added and the changes we performed to support them. Both abstract syntax tree processor and interpreter were updated to include these language constructs, since the former builds models and the latter executes the constructs.

\paragraph{Print statements} Many introductory programming assignments use printing to console to verify whether a program is correct or incorrect. The verification of correctness determines whether a program should be repaired or not. Similar to other programming languages, computations in Python can be performed using variables or inside the parentheses of a print statement, removing the need for variables altogether. Hence, while evaluating the similarity between two programs or performing a repair, it is crucial to match the contents inside these print statements. CLARA's authors did incorporate the parsing of print statements. However, when Python 3.x was introduced, the print operation switched from a statement to a function call. Furthermore, when CLARA's authors updated the implementation to work with Python 3.x, not all of the code was updated, leading to the loss of functionality of the print operation. We updated the section of the abstract syntax tree processor that checks for function calls by checking if the $\kw{print}$ function was called, and updated the model accordingly. Once the $\kw{print}$ function was correctly processed, the comparison of the print operation during the matching and repair processes was handled automatically. 

\newsavebox{\printCorrectListing}
\begin{lrbox}{\printCorrectListing}
\begin{minipage}[b]{0.40\textwidth}
\begin{python}
def isEven(val):
    print(val 
\end{python}
\end{minipage}
\end{lrbox}

\newsavebox{\printIncorrectListing}
\begin{lrbox}{\printIncorrectListing}
\begin{minipage}[b]{0.40\textwidth}
\begin{python}
def isEven(val):
    print(val 
\end{python}
\end{minipage}
\end{lrbox}

\paragraph{Import statements} Certain introductory programming assignments require the use of external libraries, such as $\kw{math}$, $\kw{re}$ or $\kw{string}$. These libraries provide access to functions like $\kw{sqrt}$, $\kw{ceil}$, $\kw{search}$ (regular expressions), or $\kw{format}$ (a string). Hence, for CLARA to execute these functions while performing a repair or a match, it is essential to have the functionality to parse and record the data within these import statements. The original implementation of CLARA ignored import statements, causing the program to crash during the repair or alignment processes, as their corresponding functions cannot be found while executing the function trace.

\newsavebox{\impStmt}
\begin{lrbox}{\impStmt}
\begin{minipage}[b]{0.45\textwidth}
\lstset{numbers=left,stepnumber=1,escapechar=;,}
\begin{python}
import math;\label{line:regular-import};
from math import sqrt;\label{line:function-import};
from math import *;\label{line:wildcard-import};
import math as m;\label{line:alias-lib-import};
from math import sqrt as sq;\label{line:alias-func-import};
\end{python}
\end{minipage}
\end{lrbox}

\newsavebox{\impDict}
\begin{lrbox}{\impDict}
\begin{minipage}[b]{0.33\textwidth}
\lstset{numbers=left,stepnumber=1,escapechar=;,}
\begin{python}
{math: math}
{sqrt:[sqrt, math]}
{*: [*, math]}
{m: math}
{sq: [sqrt, math]}
\end{python}
\end{minipage}
\end{lrbox}


Since CLARA has its own version of an abstract syntax tree processor and interpreter for Python, after parsing and processing import statements, we represent and store them in a way such that a function call is successfully recognized during execution by the interpreter. We created a section in the abstract syntax tree processor to deal with import statements and stored them in a nested global dictionary, which is provided to the interpreter to be accessed during execution. 

\paragraph{Variable assignment} In Python 3.x, a programmer can use a variable assignment based on list deconstruction or unpacking. For instance, the statement $\kw{a, b, c = [1, 2, 3]}$ is convenient to assign the values 1, 2, and 3 to variables $\kw{a}$, $\kw{b}$ and $\kw{c}$, respectively. These types of assignments are commonly used in introductory programming assignments. We updated the abstract syntax tree processor to recognize and process multiple assignments from a single statement. We separated the assignments with their corresponding expressions and added each assignment as an individual expression to the model. Without these changes, CLARA's original implementation produces an error stating that multiple assignments within a single line are not supported, halting the repair and alignment processes.

\paragraph{Built-in Python functions} CLARA's interpreter helps recognize functions in the model and execute them in Python.
Therefore, common built-in functions like $\kw{max}$, $\kw{sum}$ or $\kw{len}$ are individually defined in the interpreter using auxiliary functions like $\kw{execute\_max}$, $\kw{execute\_sum}$ or $\kw{execute\_len}$, respectively. These auxiliary functions implement the expected functionality. Since Python has a substantial collection of built-in functions, manual addition of every function was not included in the interpreter, causing CLARA to fail during the alignment and repair processes. Our approach to circumvent the need for manual addition of every function is to use Python's internal dictionary named $\kw{builtins}$. Every time a function is called, we verify whether it is a built-in Python function using such dictionary. If this is the case, we proceed to execute the function. As a result, all auxiliary functions of the type $\kw{execute\_XYZ}$ are not needed anymore.

\newsavebox{\varCorrect}
\begin{lrbox}{\varCorrect}
\begin{minipage}[b]{0.30\textwidth}
\begin{python}
def cap():
    a = [1, 2, 3]
    s = 0
    for x in a:
        s += a
    f = 2
    return a
\end{python}
\end{minipage}
\end{lrbox}

\newsavebox{\varInCorrect}
\begin{lrbox}{\varInCorrect}
\begin{minipage}[b]{0.30\textwidth}
\begin{python}
def cap():
    a = [1, 2, 3]
    s = 0
    for x in a:
        s += a
    return a
\end{python}
\end{minipage}
\end{lrbox}

\newsavebox{\varFeedback}
\begin{lrbox}{\varFeedback}
\begin{minipage}[b]{0.9\textwidth}
\begin{lstlisting}[numbers=none,keywordstyle=\color{black},stringstyle=\color{black}, commentstyle=\color{black},xleftmargin=1pt, columns=fullflexible]
1) Add 'new_f := new_f' at the beginning of function 'cap' (cost=1.0)
2) Add 'new_f := 2' *after* the 'for' loop (cost=1.0)
\end{lstlisting}
\end{minipage}
\end{lrbox}


\paragraph{Variable additions and deletions} CLARA expects all variable declarations to have a definition in the first location of the program. In other words, the beginning of the program contains assignments for every variable, and the rest of the program makes use of those variables. Hence, a new variable cannot be declared later in the program. If this happens, CLARA outputs unnecessary repairs and the final mapping of variables may be inaccurate. 
We modified the abstract syntax tree processor by removing the restriction of requiring variables to be declared in the first location, that is, new variables can be declared at any point in the program. Additionally, we modified the list of repairs generated by CLARA such that repairs suggesting to create and assign the same variable are no longer output.

\newsavebox{\mapCorrect}
\begin{lrbox}{\mapCorrect}
\begin{minipage}[b]{0.30\textwidth}
\begin{python}
def cap():
    a = [1, 2, 3]
    s = 0
    for x in a:
    s += a
    return a
\end{python}
\end{minipage}
\end{lrbox}
\newsavebox{\mapInCorrect}

\begin{lrbox}{\mapInCorrect}
\begin{minipage}[b]{0.30\textwidth}
\begin{python}
def cap():
    a = [1, 2, 3]
    s = 0
    for x in a:
        s += a
    f = 2
    g = 3
    return a
\end{python}
\end{minipage}
\end{lrbox}

\newsavebox{\mapFeedback}
\begin{lrbox}{\mapFeedback}
\begin{minipage}[b]{.9\textwidth}
\begin{lstlisting}[numbers=none,keywordstyle=\color{black},stringstyle=\color{black}, commentstyle=\color{black},xleftmargin=1pt, columns=fullflexible]
1) Delete 'g := 3' *after* the 'for' loop (cost=1.0)
\end{lstlisting}
\end{minipage}
\end{lrbox}

\begin{figure}
\centering
\subfloat[Correct program\label{fig:mapCorrect}]{\usebox{\mapCorrect}}\qquad\qquad
\subfloat[Incorrect program\label{fig:mapIncorrect}]{\usebox{\mapInCorrect}}\\
\subfloat[Suggested repair\label{fig:mapFeedback}]{\usebox{\mapFeedback}}
\caption{Correct and incorrect programs and the corresponding suggested repair that indicates to delete a single variable rather than two variables ($\kw{g}$ and $\kw{f}$)}
\label{fig:mapWrong}
\end{figure}

During the repair process, CLARA creates a mapping of variables from the correct program to an incorrect program. This mapping is based on the variable tracing performed throughout the process. A dictionary is used to store the mapping. It uses variables from the correct program as keys and incorrect program variables as values. While updating the source code to include our modifications, we noticed that, if more than one extra variable is declared, CLARA does not suggest deleting more than one variable, resulting in an incorrect final variable mapping. Figure~\ref{fig:mapWrong} illustrates this issue with two programs such that the incorrect program contains two extra variables, $\kw{g}$ and $\kw{f}$, that must be deleted; however, the suggested repair does not mention $\kw{f}$. The internal dictionary is as follows: $\kw{\{ a : a, s : 0, x : x, - : g\}}$, where variable $\kw{f}$ has been omitted. In this dictionary, variable addition and deletion are represented by the $\kw{*}$ and $\kw{-}$ keys, respectively. Since it is a dictionary, one of the deletions is overwritten as the key is the same. Note that, if the variables $\kw{f}$ and $\kw{g}$ were in the correct program, CLARA would suggest to add two new variables. This never results in an incorrect mapping problem because, in the dictionary, they are represented as $\kw{\{ f : * , g : * \}}$. On the contrary, deletions do not work as expected since $\kw{\{ - : g, - : f \}}$ is not allowed and, therefore, we only get the suggestion to remove one variable. We thus adjusted the internal dictionary to save an array of values in the case of deletions, i.e., $\kw{\{ - : \langle f, g \rangle \}}$ in our example.


\paragraph{Input statements} One of the commonalities of introductory programming assignments is that they evaluate the correctness of a program based on test cases using console input and output. The original implementation of CLARA, however, does not support standard input. All inputs have to be provided via the command line as function arguments. This is not always possible since input arguments can be multiple lines long and do not have the same length. Therefore, we updated CLARA to read all of the inputs using an argument file and store them in an internal list accessible by the interpreter. We updated the interpreter to handle calls to the $\kw{input}$ function separately. So a call like $\kw{x = input()}$ is handled as follows: we extract the first element from the internal list and assign it to variable $\kw{x}$. If there are subsequent calls to $\kw{input}$, we keep extracting elements from the internal list. As a result, this modification allows us to repair programs that had inputs of different length.

However, while adding this feature, we encountered another problem. Since CLARA nests the expressions of variables while creating its model, it creates copies of the $\kw{input}$ function when there should only be a single call. For example, consider the following Python statement:

\begin{center}
\begin{minipage}{0.42\textwidth}
\begin{python}
a, b, c = input().split()
\end{python}
\end{minipage}
\end{center}

It becomes the following statements in the model:

\begin{center}
\begin{minipage}{0.5\textwidth}
\begin{lstlisting}[numbers=none,keywordstyle=\color{black},stringstyle=\color{black}, commentstyle=\color{black},xleftmargin=1pt, columns=fullflexible]
a = GetElement(split(input()), 0)
b = GetElement(split(input()), 1)
c = GetElement(split(input()), 2)
\end{lstlisting}
\end{minipage}
\end{center}

This change results in $\kw{input}$ being called three times, where it should have been called just once. Since it is possible for this problem to occur with other function calls as well, we updated the abstract syntax tree processor to create a new variable to store the result of calling the $\kw{input}$ function. Additionally, the expression referencing the function references the new variable instead. Therefore, using the above example, the statements in the model are as follows:

\begin{center}
\begin{minipage}{0.5\textwidth}
\begin{lstlisting}[numbers=none,keywordstyle=\color{black},stringstyle=\color{black}, commentstyle=\color{black},xleftmargin=1pt, columns=fullflexible]
input_val = input()
a = GetElement(split(input_val), 0)
b = GetElement(split(input_val), 1)
c = GetElement(split(input_val), 2)
\end{lstlisting}
\end{minipage}
\end{center}

Repetition of expressions is expected if we have multiple variable declarations in a single line during model creation. If these expressions include side-effecting functions, we can have a similar problem as we had with the $\kw{input}$ function. It is challenging to automatically detect whether a function is side-effecting, and creating new variables for every single function call in a program is also challenging to handle due to the addition of multiple variables. Furthermore, it can cause a mismatch in the number of variables between the correct program and the incorrect program, resulting in unexpected repairs. However, this is not a problem for the $\kw{input}$ function, as we expect both the programs to have the same number of calls to the $\kw{input}$ function. We adjusted the source code to address this issue to receive an optional list of the side-effecting functions via command line. New variables will be created for each of these functions similar to $\kw{input}$, aiding us in solving the problem and limiting the addition of extra variables. In practice, one can apply a preprocessing step detecting side-effecting functions and add them to this command-line list.

\section{Alignment and repair modifications}
\label{sec:mod-repair}

Before performing the modifications described above, CLARA's original implementation did not output any model when processing a program containing any unsupported statements. Since both the alignment and repair processes depend on models, both processes were thus not executed. After performing the modifications described above, the alignment and repair processes worked properly for many programs. However, some programs still failed. In this section, we report our modifications to the alignment and repair processes of CLARA's original implementation. Note that these modifications were necessary because of the modifications made to the parser and interpreter presented above.

\paragraph{Nested functions} The original implementation of CLARA is not able to parse nested functions as functions cannot store other functions in the model. A function is thus only allowed to store expressions. We updated the model to support nested functions by creating a link between two or more functions. After adding it to the model, we updated the processes to align and repair these functions successfully. Both processes involve creating a one-to-one mapping between variables. Therefore, we need to ensure that the variables inside the nested functions are not involved in the mapping as those exist in a different environment. We treat each nested function as a variable, which is evaluated during trace execution, and its return value is substituted by the variable calling the function. If the function is called on its own and does not have a return value, we check if it is printing to standard output. If that is the case, the expression being printed is added to the standard output of the outer function. Since CLARA performs variable tracing and tracks the values a variable holds throughout the program, CLARA places more importance on the variable's values than its expression. This helps eliminate the need for function inlining. Our aim while updating the alignment and repair processes was to avoid the suggestion of creating/deleting nested functions if the other program is performing the exact computation without using nested functions.

\paragraph{Applying repairs} CLARA's repair process is sound and complete for the test case provided as input~\cite{DBLP:conf/pldi/GulwaniRZ18}. However, it is typically the case that introductory programming assignments are evaluated with a variety of test cases. Therefore, we aim to determine whether the incorrect program provided as input is repaired for only that particular test case or for all test cases available. To accomplish this, we need to convert CLARA's output into actual repairs and, then, apply these repairs to the model of the program. Note that the repair process focuses solely on models and not the original source code; therefore, we decided to repair the program's model rather than the source code.

Every statement output by CLARA contains the variables involved in the repair for both the correct and incorrect programs, the associated location in the correct program, and the associated expression from the correct program. Hence, we must extract the rest of the necessary information, i.e., the location in the incorrect program and the expression. Every function in a program groups and stores its variables and their associated expressions by location. There are three types of repairs: variable deletion, variable addition, and changing the variable definition. In the case of variable deletion, we access its corresponding location expressions and remove the variable from the list. Similarly, we add the variable and the new expression to its corresponding location expressions for variable additions. Finally, we substitute the variable's expression in its corresponding location for variable changes.

However, CLARA's output is not sorted; therefore, for variable additions and changes, we must ensure expressions are added in such an order that the variables being used exist. For example, if we have the following output:

\begin{lstlisting}[numbers=none,keywordstyle=\color{black},stringstyle=\color{black}, commentstyle=\color{black},xleftmargin=1pt, columns=fullflexible]
1) Change 'a = x + 5' to 'a = m + 5'
2) Add 'm = 3'
\end{lstlisting}

\newsavebox{\AppRepProg}
\begin{lrbox}{\AppRepProg}
\begin{minipage}[b]{0.35\textwidth}
\begin{python}
a = 3
b = a + 1
for x in range(0, 2):
    b += x
b += a
\end{python}
\end{minipage}
\end{lrbox}

\newsavebox{\AppRepModel}
\begin{lrbox}{\AppRepModel}
\begin{lstlisting}[numbers=none,keywordstyle=\color{black},stringstyle=\color{black}, commentstyle=\color{black},xleftmargin=0.1\textwidth, columns=fullflexible]
Loc 1 (around the beginning of function)
-----------------------------------
  a := 3
  b := Add(a', 1)
  iter#0 := range(0, 2)
  ind#0 := 0
-----------------------------------
  True -> 2   False -> None

Loc 2 (the condition of the 'for' loop at line 3)
-----------------------------------
  $cond := Lt(ind#0, len(iter#0))
-----------------------------------
  True -> 4   False -> 3

Loc 3 (*after* the 'for' loop starting at line 3)
-----------------------------------
  b := AssAdd(b, a)
-----------------------------------
  True -> None   False -> None

Loc 4 (inside the body of the 'for' loop beginning at line 4)
-----------------------------------
  x := GetElement(iter#0, ind#0)
  ind#0 := Add(ind#0, 1)
  b := AssAdd(b, x')
-----------------------------------
  True -> 2   False -> None
\end{lstlisting}
\end{lrbox}

\begin{figure}
\centering
\usebox{\AppRepProg}
\caption{Sample program to illustrate how CLARA differentiates between variable definition and usage}
\label{fig:AppRepProg}
\end{figure}

\begin{figure}
\centering
\usebox{\AppRepModel}
\caption{Model of the program in Figure~\ref{fig:AppRepProg} that differentiates between variable definition and usage, e.g., $\kw{a}$ and $\kw{a'}$}
\label{fig:AppRepModel}
\end{figure}

We must ensure that $\kw{m}$ is defined before $\kw{a}$; otherwise, an error is thrown during program execution. As mentioned earlier, CLARA differentiates between variable definition and variable usage. An example of this can be seen in Figures~\ref{fig:AppRepProg} and~\ref{fig:AppRepModel}, where location 1 defines variable $\kw{a}$ and uses it, denoted as $\kw{a'}$. Note that location 3 uses the same variable; however, in this case, location~3 does not (re)define variable $\kw{a}$; therefore, it does not use $\kw{a'}$. As a result, we need to deal with an additional problem involving variable definition. For variable usage, we must ensure that variables exist and are defined earlier in the location. If we add a new variable definition within a location, we need to make sure that usages of that variable are updated. For instance, if we add a definition of $\kw{a}$ into location 3, we must change $\kw{a}$ to $\kw{a'}$ when it is used to update $\kw{b}$. Therefore, every time a new variable is added or deleted, we execute a trace of the function to help us determine which variables we have access to before adding a repair. If all of the variables used in that expression are defined, we add the repair. Otherwise, we continue adding the rest of the repairs and come back to the ones we skipped earlier until no more repairs are available.

Once all the repairs have been applied, we rerun the repair process to determine if any additional repairs are suggested for the same test case. If this happens, we halt and conclude that the program is not successfully repaired. Otherwise, we conclude that the incorrect program is repaired for that particular test case, run the repair process for all other available test cases, and record if any repairs are suggested. If no other repairs are suggested overall, we can successfully conclude that the incorrect program has been fully repaired for all test cases.

\section{Flexible program alignment}
\label{sec:flex-align}

One of CLARA's main limitations is that it requires both the correct and the incorrect programs to have similar control flows in order to proceed with the repair process. It is uncommon to find many programs containing the same control flow, which reduces the number of programs CLARA can repair in practice.~\citet{claraRefactor} reported that CLARA's repair process did not work for 35.5\% of incorrect programs using several introductory programming assignments. Our experiments below also confirm these findings.

We propose an algorithm that creates a flexible alignment between the control flow graphs of the correct and incorrect programs. This flexible alignment takes into consideration both semantic and topological information of the graphs. Our main goal is to reduce mismatches in the alignment process; however, this comes with a penalty: since it is approximate, it is possible to obtain an alignment that makes the repair process fail. We discuss these issues in this section.

\newsavebox{\GMCorrect}
\begin{lrbox}{\GMCorrect}
\begin{minipage}[b]{0.40\textwidth}
\begin{python}
def f(a):
    x, m, s = 0, 101, 0
    while x < len(a):
        s += a[x]
        if m > a[x]:
            m = a[x]
        x += 1
    print(s + "," + m)
\end{python}
\end{minipage}
\end{lrbox}

\newsavebox{\GMIncorrect}
\begin{lrbox}{\GMIncorrect}
\begin{minipage}[b]{0.40\textwidth}
\begin{python}
def g(a):
    i, m, s = 0, 101, 0
    while i < len(a):
        s += a[i]
        if m < a[i]:
            m = a[i]
        i += 1
        if m == 0:
            i -= 1
    print(s + "," + m)
\end{python}
\end{minipage}
\end{lrbox}

\begin{figure}
\centering
\subfloat[Correct Program\label{fig:CorrectProgGM}]{\usebox{\GMCorrect}}\qquad\qquad
\subfloat[Incorrect Program\label{fig:IncorrectProgGM}]{\usebox{\GMIncorrect}}
\caption{Correct and incorrect programs used to illustrate flexible alignment}
\label{fig:corr&inCorrForGraphMatch}
\end{figure}

Figure~\ref{fig:corr&inCorrForGraphMatch} presents a simplified version of the programs presented in Section~\ref{sec:overview}. We use these two programs to illustrate our discussions in this section. Note that, in this section, we assume CLARA's optimization of replacing if statements containing no loops by ternary operators is disabled.

\subsection{Creation of the control flow graph}

Each model created by CLARA is based on the program's control flow. Hence, we decided to utilize the information available to us in the model to create control flow graphs for both the correct and incorrect programs. Each location is a node in the control flow graph, and the location's transitions are the edges in the graph. Since many transitions pointed to $\kw{None}$, indicating that no transition exists, we decided to create a special node for $\kw{None}$. 

Each node in the control flow graph contains a location number, the corresponding location's expressions, the starting line number of that location, a description, and a multiset of labels. This multiset contains all the semantic information we can extract from the (nested) expressions of the location at hand. Each expression contains semantic information in terms of constants and operation semantics. Operation semantics allow us to identify the type of statement, e.g., addition or subtraction. Unlike expressions, which can hold different values due to variables, we want the elements of a label to always remain constant for a particular expression. Therefore, variable names are not added to the multiset of labels. For example, consider the following Python statements:

\begin{center}
\begin{minipage}[b]{0.22\textwidth}
\begin{python}
a = [1, 5, 6]
b = a[0]
c = b + 1
\end{python}
\end{minipage}
\end{center}

We extract a multiset of labels for each statement. Note that these labels correspond to expressions in the model derived from the Python statements. Since $\kw{a}$ is a list formed by the elements 1, 5 and 6, it contains the labels $\kw{\{ListInit, 1, 5, 6\}}$. Also, $\kw{b}$ is initialized with the first element of $\kw{a}$, so it has the labels $\kw{\{GetElement, 0\}}$; similarly, $\kw{c}$ is initialized with $\kw{b + 1}$, so it has the labels, $\kw{\{Add, 1\}}$. Therefore, the labels assigned to the node is the union of all of these multisets, i.e, $\kw{\{ListInit, 1, 5, 6, GetElement, 0, Add, 1\}}$ (note that 1 appears twice). Recall that we exclude the variable names $\kw{a}$, $\kw{b}$ and $\kw{c}$ from the labels of the node as the variables do not necessarily evaluate to the same value at every point in the program.

\begin{figure}
\centering
\subfloat[Correct program\label{fig:CorrectCFG}]{\includegraphics[scale=0.75]{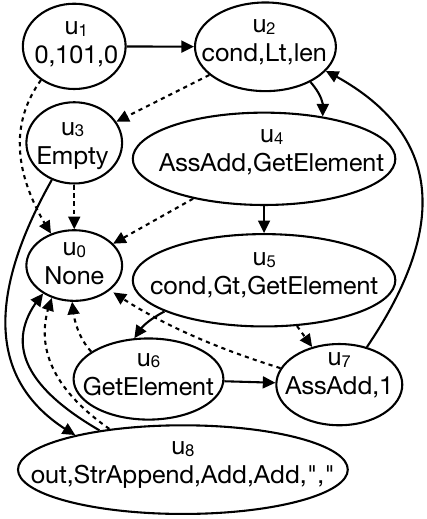}}\qquad\qquad
\subfloat[Incorrect program\label{fig:IncorrectCFG}]{\includegraphics[scale=0.75]{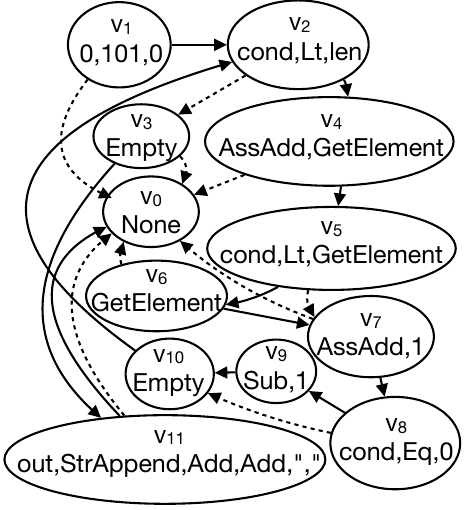}}
\caption{Control flow graphs derived from the programs in Figure~\ref{fig:corr&inCorrForGraphMatch}}
\label{fig:CFGs}
\end{figure}

We iterate through the model of a function to create the nodes  of the control flow graph with the information mentioned above. After the nodes have been created, we connect the nodes with edges representing the transitions between the model locations. Each edge is annotated with a True or False label based on the type of the transition.
Figure~\ref{fig:CFGs} shows the control flow graphs obtained from the programs in Figure~\ref{fig:corr&inCorrForGraphMatch}. Each node in the figure contains the location number and its corresponding multiset of labels. The expressions, line numbers, and descriptions have been omitted from the figure to improve readability. The dotted arrows represent False transitions, and the solid arrows represent True transitions. The node with the label $\kw{Empty}$ represents a location in the program model that does not contain any expressions, which corresponds to the end of the loop.

\subsection{Alignment algorithm}

\begin{algorithm}
\caption{FlexAlign}
\label{alg:GM}
\SetKwInOut{Input}{input}
\SetKwInOut{Output}{input/output}
\Input{$G_C = (U, E)$ control flow graph of the correct program; $G_I = (V, F)$ control flow graph of the incorrect program}
\Output{$\phi_{best}: U \rightarrow V$}
// Initialize best alignment and similarity. \\
$\phi_{best} \gets \{\}, s_{best} \gets 0$ \\
// Check every permutation. \\
\For{$\phi \in Permutations(U, V)$}{ \label{alg-line:permutations}
    // $s$ is the similarity of $\phi$. \\
    $s \gets 0$ \\
    \For{$u \in dom~\phi$}{ \label{alg-line:u-in-phi}
        $v \gets \phi(u)$ \\ \label{alg-line:v-in-phi}
        // Jaccard similarity between the labels of $u$ and $v$. \\
        $s_{label} \gets Jaccard(L(u), L(v))$ \\ \label{alg-line:jaccard}
        // $u'$ and $u''$ ($v'$ and $v''$) are the neighbors of $u$ ($v$). \\
        Let $\{u \xrightarrow{True} u', u \xrightarrow{False} u''\} \subseteq E$, $\{v \xrightarrow{True} v', v \xrightarrow{False} v''\} \subseteq F$ \\ \label{alg-line:edge-sim-1}
        // Edge similarity is 0.5 by default. \\
        $s_{edge} \gets 0.5$ \\
        // Check if the neighbors match. \\
        \uIf{$v' = \phi(u') \land v'' = \phi(u'')$}{
            $s_{edge} \gets 1$
        }\ElseIf{$v' \neq \phi(u') \land v'' \neq \phi(u'')$}{
            $s_{edge} \gets 0$ \label{alg-line:edge-sim-2}
        }
        // Aggregate both similarities (same importance). \\
        $s \gets s + (s_{label} + s_{edge})/2$ \label{alg-line:update-similarity}
    }
    // Update the best alignment found. \\
    \If{$s > s_{best}$} { \label{alg-line:update-best-1}
        $s_{best} \gets s$, $\phi_{best} \gets \phi$ \label{alg-line:update-best-2}
    }
}
\end{algorithm}

Our graph alignment process aims to find a mapping between the nodes of the control flow graphs of the correct and incorrect programs. Algorithm~\ref{alg:GM} aims to compute such mapping taking into consideration both the semantic and topological information of the graph, i.e., labels and edges, respectively. The algorithms takes the two control flow graphs of the incorrect and correct programs as input, denoted as $G_C = (U, E)$ and $G_I = (V, F)$, respectively, where $U$ and $V$ are sets of nodes, and $E$ and $F$ are sets of edges. The algorithm aims to find a mapping $\phi_{best}$ from $U$ to $V$. To accomplish this, it explores all possible permutations of these mappings, that is, all combinations of nodes in $U$ mapped to nodes in $V$, which is performed by the $Permutations$ function (line~\ref{alg-line:permutations}). Let $\phi$ be one of these permutations. The similarity of $\phi$, denoted by $s$, is computed by considering label and edge similarities. For each node $u$ present in $\phi$ (line~\ref{alg-line:u-in-phi}) and its corresponding $v$ (line~\ref{alg-line:v-in-phi}), the algorithm computes the Jaccard similarity between the multisets of labels of both nodes (line~\ref{alg-line:jaccard}). The formula for the multisets $L(u)$ and $L(v)$ is as follows~\cite{DBLP:journals/corr/abs-2110-09619}:

\begin{displaymath}
Jaccard(L(u), L(v)) = \frac{\sum_{x \in L(u) \cap L(v)} min(W(x, L(u)), W(x, L(v)))}{\sum_{x \in L(u) \cup L(v)} max(W(x, L(u)), W(x, L(v)))}
\end{displaymath}

\begin{table}
\centering
\caption{Multisets of labels and their corresponding Jaccard similarities}
\begin{tabular}[t]{|c|c|c|}
\hline
\textbf{Multiset $L(u)$} & \textbf{Multiset $L(v)$}
 & \textbf{Jaccard} \\
\hline\hline
GetElement, 0, 0 & GetElement, 0 & 0.667\\
\hline
GetElement, 0 & GetElement, 0 & 1\\
\hline
cond, Lt, ind\#0, len, iter\#0 & cond, Gt, ind\#1, len, iter\#1 & 0.25\\
\hline
ListInit, 5, 6 & ListInit, 8, 9 & 0.2\\
\hline
\end{tabular}
\label{fig:label&jaccard}
\end{table}

\noindent where $W(x, S)$ is the number of times $x$ appears in the multiset $S$. Note that $0 \le Jaccard(L(u), L(v)) \le 1$, where zero indicates no similarity and one indicates both multisets are equal. Table~\ref{fig:label&jaccard} presents several examples of multisets of labels and their corresponding Jaccard similarities. We rely on Jaccard similarity because it penalizes dissimilarities between the multisets compared, intersection divided by union, more than others like Sørensen–Dice.

Once Jaccard similarity is computed, the algorithm computes edge similarity (lines~\ref{alg-line:edge-sim-1}--\ref{alg-line:edge-sim-2}). It considers both neighbors of $u$ and $v$ that correspond to the False and True transitions in each control flow graph. If both sets of neighbors are mapped in $\phi$, the edge similarity is one; if none of them are mapped in $\phi$, the edge similarity is zero; otherwise, the edge similarity is 0.5 (one but not the other is mapped).

Both similarities are combined using the same importance and added to the current similarity of $\phi$ (line~\ref{alg-line:update-similarity}). Finally, the graph alignment that is output is the one with highest similarity (lines~\ref{alg-line:update-best-1}--\ref{alg-line:update-best-2}).

Note that, even though we expect to deal with small programs, the number of nodes in a program in an introductory programming assignment typically ranges between 10 and 20; therefore, it is possible to have more than 3 million permutations. In practice, we aim to find permutations with high similarities first. To accomplish this, we use a heuristic in which we explore the permutations in ascending order by semantic similarity. Furthermore, we only explore the top-$k$ permutations found using this approach.

\begin{figure}
\centering
\includegraphics[scale=0.80]{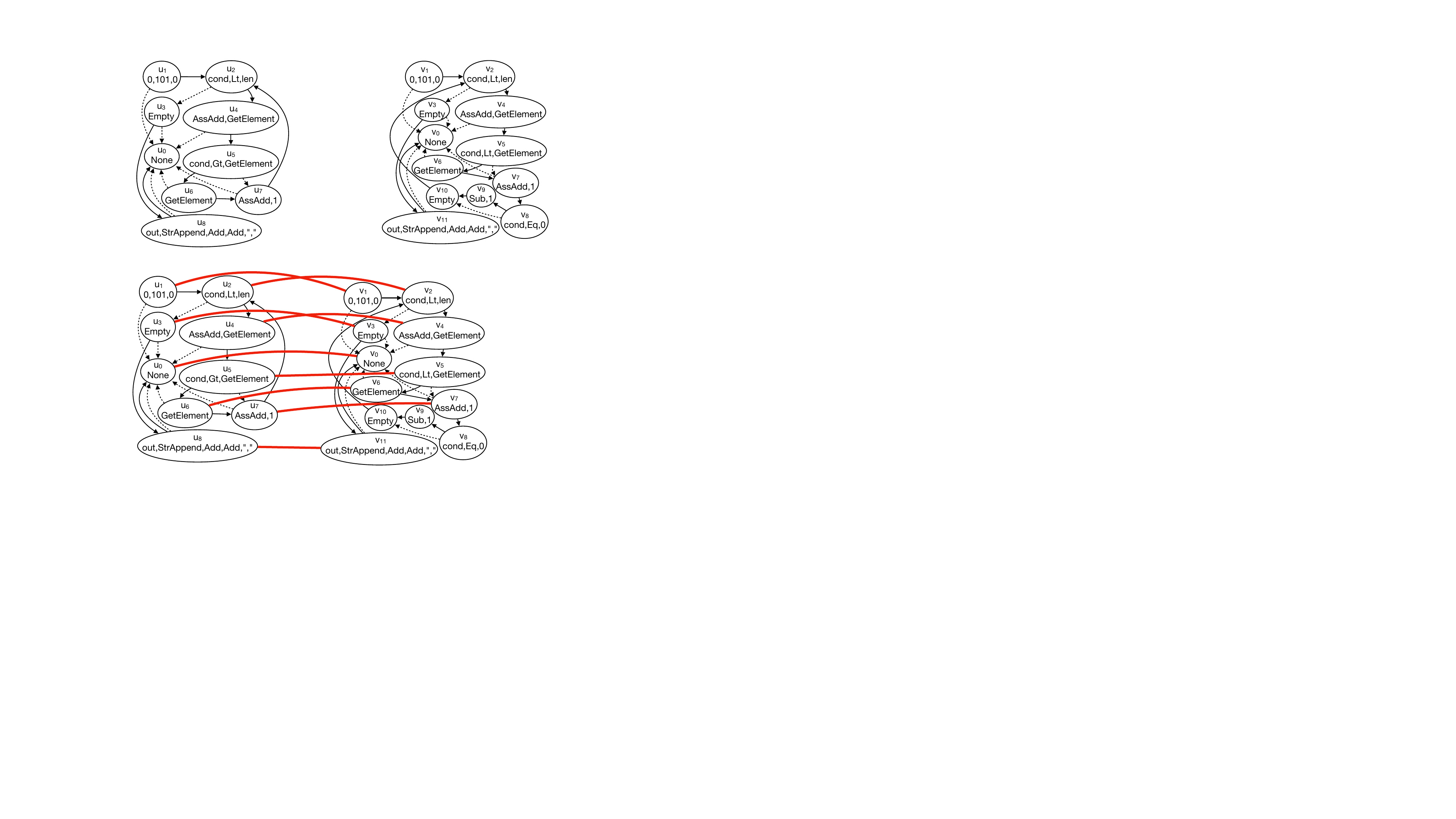}
\caption{Alignment between the control flow graphs from Figure~\ref{fig:CFGs}}
\label{fig:matchCFGs}
\end{figure}

Figure~\ref{fig:matchCFGs} presents an alignment between the control flow graphs discussed above. The control flow graph of the correct program is on the left side, and the graph of the incorrect program is on the right side. The solid red lines indicate the mapping $\phi$ between the nodes. As it can be observed, nodes $v_8$, $v_9$ and $v_{10}$ are not mapped to any nodes in the correct program. These locations correspond to the second $\kw{if}$ statement in the incorrect program.

Algorithm~\ref{alg:GM} always produces an alignment between the input programs. One can use $s$, the similarity of an alignment, to decide whether or not to proceed with the repair process. In our experiments below, we use a threshold over $s$ to proceed to repair the programs.

\begin{algorithm}
\caption{RecreateModel}
\label{alg:Recreat}

\SetKwInOut{Input}{input}
\SetKwInOut{Output}{output}
\Input{$G_C = (U, E)$ control flow graph of the correct program; $G_I = (V, F)$ control flow graph of the incorrect program; $\phi: U \rightarrow V$ alignment}
\Output{$M_I = (W, J)$ recreated model of the incorrect program}

// Initialize $M_I$ with the nodes in $G_I$ and empty edges. \\
$W \gets V, J \gets \emptyset$ \\

// Correct is smaller than incorrect; remove extra nodes. \\
\If{$|U| < |V|$}{ \label{alg-line:smaller-1}
    // If $v$ is not in $\phi$, remove from $W$. \\
    \For{$v \in V$ {\normalfont such that} $v \notin ran~\phi$}{
        $W \gets W \setminus \{v\}$ \label{alg-line:smaller-2}
    }
}
// Incorrect is smaller than correct; add extra nodes. \\
\If{$|U| > |V|$}{ \label{alg-line:greater-1}
    // If $u$ is not in $\phi$, add new node to $W$ and $\phi$. \\
    \For{$u \in U$ {\normalfont such that} $u \notin dom~\phi$}{
        $v' \gets CreateNode()$ \\
        $W \gets W \cup \{v'\}$ \\
        $\phi(u) \gets v'$ \label{alg-line:greater-2}
    }
}
// Update $J$ to reflect the edges in $G_C$. \\
\For{$u \in dom~\phi$}{\label{alg-line:edge-update-1}
    $v \gets \phi(u)$ \\
    Let $\{u \xrightarrow{True} u', u \xrightarrow{False} u''\} \subseteq E$ \\
    $J \gets J \cup \{v \xrightarrow{True} \phi(u'), v \xrightarrow{False} \phi(u'')\}$\label{alg-line:edge-update-2}
}
\end{algorithm}

\subsection{Recreating the model}

Once we have computed an alignment $\phi$ between the control flow graphs of the correct and the incorrect programs, we have to recreate the input that CLARA's repair process requires. Since the process receives models for each program, we rely on $\phi$ and the model of the correct program to recreate the new model. There are three possibilities. First, the control flow graph of the correct program has less nodes than the graph of the incorrect program. In this case, nodes need to be removed from the new model. Second, the control flow graph of the correct program has more nodes, which implies that new nodes need to be added to the new model. Third, both graphs have the same size, so no nodes need to be added or removed. In all of the three cases, edges or nodes might need rearrangement in the new model after an alignment has been determined.

Algorithm~\ref{alg:Recreat} takes as input both control flow graphs ($G_C$ and $G_I$) as well as the alignment between them ($\phi$). It outputs $M_I$, the recreated model of the incorrect program. If there are more nodes in the incorrect program than in the correct program, the algorithm removes the extra nodes from the new model (lines~\ref{alg-line:smaller-1}--\ref{alg-line:smaller-2}). Note that, for every node that is deleted, its corresponding expressions and edges (transitions) are deleted too. If there are less nodes, it generates new, fresh nodes using the $CreateNode$ function, which are added to both the new model and $\phi$ (lines~\ref{alg-line:greater-1}--\ref{alg-line:greater-2}). Note that these new nodes do not contain any expressions because we do not know yet the corresponding variables associated with those expressions. During the repair process, additions of new expressions will be suggested for these new nodes. Finally, the algorithm updates the edges of the new model based on the edges of the correct program (lines~\ref{alg-line:edge-update-1}--\ref{alg-line:edge-update-2}).


\section{Evaluation}
\label{sec:evaluation}

To evaluate the performance of CLARA and the improvements of our flexible alignment scheme, we built a dataset of correct and incorrect programs from the online programming website Codeforces (\url{https://codeforces.com}). It is an online platform that hosts competitive programming contests and programming problems divided into multiple difficulty ratings.
Submissions by users, both correct and incorrect, are publicly available.
To evaluate the correctness of a program, the platform executes it on several test cases.
Codeforces programming problems follow the same structure: each test case must be read from the console by a given program, and such a test case consists of line-delimited parameters. The first line indicates the number of arguments, and the following lines include arguments as a single block of text that requires parsing before performing any computations to solve the problem at hand. A variety of methods can be used for such parsing, but the most common is Python's \texttt{input} function for reading console input. 

If a program does not pass a particular test case, the testing stops and the program is declared incorrect. The test case on which the program failed is provided by the platform. Therefore, we only have access to the first test case that is not passed for every incorrect program. Furthermore, because CLARA uses variable tracing to find repairs in the program, we only need the test case input, not the output it is supposed to produce. The platform does not display test cases that are longer than 50 lines; instead, it displays the initial 50 lines followed by ``\dots'' characters. As a result, these test cases are incomplete, inaccessible, and therefore, invalid for our purposes.

Taking these factors into account, in our experiments, we utilized valid submissions for twenty programming problems with the highest number of Python submissions, and fulfilled the condition of having at least one of the following: (1)~a loop, (2)~an $\kw{if}$ statement, (3)~a call to read from standard input, and (4)~a call to print to standard output. 

\subsection{Dataset and Experimental Setup}
\label{subsec:exp_setup}

Since CLARA analyzes an incorrect program based on correct programs, we selected thirty correct programs to be compared with the pool of all incorrect programs for each of the twenty programming problems. The subset of thirty correct programs for each problem was selected as follows: select the top-10 programs when sorted by date and they are from different users. Then, select the top-10 programs when sorted by size ascending and they are from different users. Repeat the same operation using descending order. All of the incorrect programs taken into account failed valid test cases as recorded by the platform. These programs were all unique and made by different users. We evaluated \jss{five} techniques as follows:

\begin{enumerate}

\item \jss{Baseline CLARA with No Alignment (CNA): The original implementation of CLARA with no modifications and no flexible alignment.}

\item SARFGEN: Sarfgen~\cite{DBLP:conf/pldi/WangSS18} is not publicly available; therefore, we simulated its alignment step. We used our proposed flexible alignment (see Algorithm~\ref{alg:GM}), considering both semantic and topological similarities (label and edge), and model recreation (see Algorithm~\ref{alg:Recreat}). However, we set a threshold: only similarities greater or equal than 0.95 are kept. This simulates Sarfgen's rigid program comparison in which both control flow graphs must perfectly match.

\item Baseline CLARA: The original implementation with the modifications described in Section~\ref{sec:modifications}. The alignment process is the one originally implemented (see Algorithm~\ref{alg:cap}).

\item FA(L): It uses our proposed flexible alignment (FA) (see Algorithm~\ref{alg:GM}) and model recreation (see Algorithm~\ref{alg:Recreat}). However, it only exploits semantic information (label) for alignments, i.e., edge similarity is always equal to one ($s_{edge} = 1$).

\item FA(L+E): Similar to FA(L) but it considers both semantic and topological similarities (label and edge) as described in Algorithm~\ref{alg:GM}.
\end{enumerate}

Note that the \jss{five} techniques rely on the same repair process, i.e., the process of CLARA's original implementation. Furthermore, we noted that the number of nodes in each program's control flow graph ranges between 1 and 70. This implies that, in many cases, we can have millions of permutations to be evaluated to compute an alignment. Therefore, for SARFGEN, FA(L) and FA(L+E), we set a limit of 1,000 permutations.  The best alignment is thus chosen from these permutations. Recall that we sort the node candidates by semantic similarity (labels) with the expectation that an alignment with a high similarity will be computed. We also set a time limit of one minute for SARFGEN, FA(L), and FA(L+E), and an overall time limit of five minutes. If any of these limits is reached, we report a timeout error. 

\begin{table*}
\centering
\caption{Program comparisons available in our dataset and summary of invalidity reasons. \jss{SGEN refers to SARFGEN while CNA refers to CLARA with no flexible alignment}.}
\label{tab:parse_errors}
\setlength\tabcolsep{4.5pt}
\begin{tabular}{|l|c|c|c|c|c|}
\hline
Reasons & \jss{CNA} & \jss{SGEN} & CLARA & FA(L) & FA(L+E)\\
\hline
Unavailable Test Cases       &\jss{25,485}   & \jss{25,498}   &   25,485 & 25,498 & 25,498   \\
Unsupported Constructs       &\jss{26,152}   & \jss{26,152}   &   26,152 & 26,152 & 26,152   \\
Unsupported Class Attributes &\jss{7,868}  & \jss{7,868}  &   7,868 & 7,868 & 7,868          \\
Keyword Arguments            &\jss{5,148}  & \jss{5,148}  &   5,148 & 5,148 & 5,148          \\
Timeout                      &\jss{9}   & \jss{172}   &   9 & 191 & 172                    \\
Total Invalid                &\jss{64,838}   & \jss{64,838}   &   64,662 & 64,857 & 64,838    \\
Total Valid                  &\jss{15,925}   & \jss{15,749}   &   15,925 & 15,730 & 15,749    \\
Total Comparisons            &\jss{80,587}   & \jss{80,587}   &   80,587 & 80,587 & 80,587    \\
\hline
\end{tabular}
\end{table*}

Table~\ref{tab:parse_errors} presents a summary of the program comparisons available in our dataset. The total number of program comparisons is 80,587, which corresponds to the Cartesian product between the total number of incorrect programs and the subset of correct programs selected as explained above. A significant portion of the comparisons was filtered out because of the following reasons: (1)~The test cases were not available,~(2)~Contained unsupported language constructs like lambda expressions or try-catch blocks,~(3)~Contained class attributes,~(4)~Contained functions with keyword arguments, and~(5)~There were timeout errors. Note that there were only 191 timeout errors in the worst case among the comparisons, which amounts to less than 1.25\% of the total valid comparisons. Therefore, timeout errors were significantly mitigated thanks to the time thresholds we established. Due to the difference in the number of timeouts for each technique, the total valid comparisons are different between them. As a result, we further filter the remaining valid comparisons to only pick the comparisons with common permutations of valid and invalid programs. This results in 15,688 common valid program comparisons available for all the techniques.

\begin{table}
\caption{Dataset statistics for valid programs, where LOC and Exprs.~respectively indicate total lines of code and number of expressions, and Diff.~is the problem's difficulty}
\centering
\begin{tabular}{|l | c| c| c| c| c| c|}
\hline
Problem & Total & Correct & Incorrect & LOC & Exprs. & Diff.\\
\hline
4A      &       1,059 &           10 & 43  &  $6.94 \pm 0.43$  &  $ 5.02 \pm 0.14$  & 800 \\
50A     &       567 &            5 &  60  &  $2.38 \pm 1.31$  &  $ 4.90 \pm 1.26$  & 800 \\
214A    &       717 &            7 &  66  &  $11.8 \pm 5.21$  &  $ 21.0 \pm 7.87$  & 800 \\
255A    &       1,856 &           12 & 96  &  $12.5 \pm 7.77$  &  $ 13.1 \pm 7.14$  & 800 \\
265A    &       300 &            6 &  25  &  $4.95 \pm 0.22$  &  $ 11.6 \pm 1.75$  & 800 \\
510A    &       1,001 &           7 &  78  &  $17.7 \pm 5.14$  &  $ 18.4 \pm 8.68$  & 800 \\
1097A   &       1,329 &           10 & 72 &   $16.1 \pm 15.6$ &   $ 17.6 \pm 13.13$ & 800 \\
1360B   &       544 &            4 &  85 &   $8.61 \pm 1.20$  &  $ 18.1 \pm 2.71$  & 800 \\
1370A   &       3,012 &           6 &  249 &  $4.53 \pm 1.09$  &  $ 10.0 \pm 3.06$  & 800 \\
1385A   &       4,779 &           6 &  428 &  $26.8 \pm 3.45$  &  $ 24.9 \pm 4.52$  & 800 \\
1391A   &       9,033 &           25 & 148 &  $5.83 \pm 2.27$  &  $ 9.23 \pm 4.35$  & 800 \\
1391B   &       318 &            6 &  51  &  $9.91 \pm 2.80$  &  $ 23.1 \pm 5.19$  & 800 \\
208A    &       447 &            5 &  65  &  $8.64 \pm 10.9$ &   $ 8.44 \pm 10.16$ & 900  \\
1A      &       877 &            5 &  125 &  $2.00 \pm 0.00$  &  $ 5.73 \pm 1.03$  & 1000 \\
1382B   &       1,532 &           5 &  145 &  $15.5 \pm 4.31$  &  $ 21.2 \pm 2.59$  & 1100 \\
492B    &       1,147 &           9 &  90  &  $9.03 \pm 8.15$  &  $ 18.7 \pm 11.41$ & 1200  \\
1363A   &       2,788 &           7 &  295 &  $17.0 \pm 9.96$ &   $ 27.5 \pm 5.96$  & 1200 \\
1364A   &       12,581 &          11 & 679 &  $8.63 \pm 5.74$  &  $ 14.5 \pm 4.65$  & 1200 \\
1369B   &       2,410 &           7 &  211 &  $8.00 \pm 5.72$  &  $ 14.4 \pm 5.63$  & 1200 \\
4C      &       1,107 &           10 & 66  &  $10.3 \pm 1.57$  &  $ 16.9 \pm 2.53$  & 1300 \\

\hline
\end{tabular}
\label{tab:correct_incorrect_num}
\end{table}

\begin{figure}
\centering
\subfloat[Lines of code]{\includegraphics[scale=0.2]{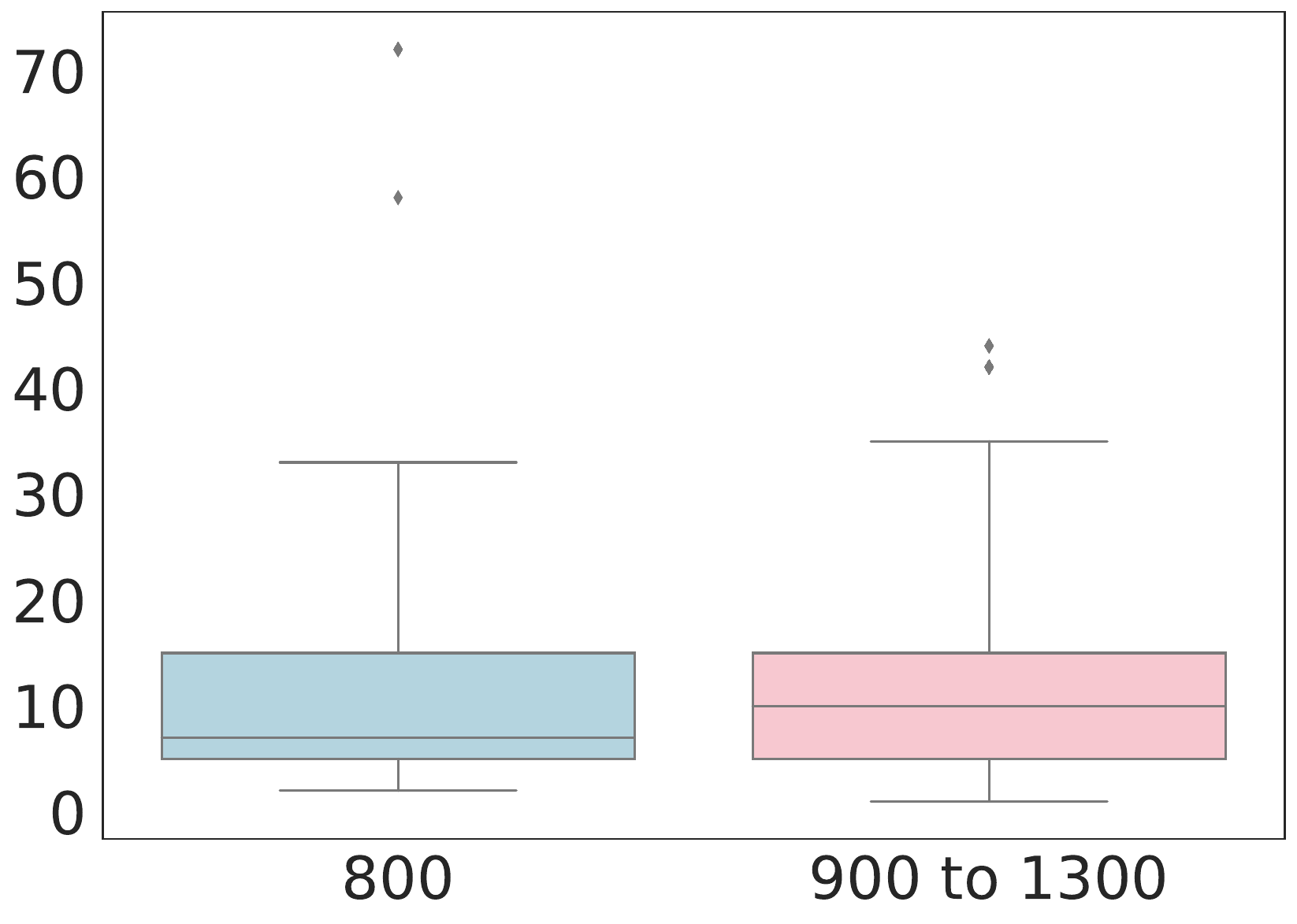}\label{subfig:locs_difficulty}}~
\subfloat[Expressions]{\includegraphics[scale=0.2]{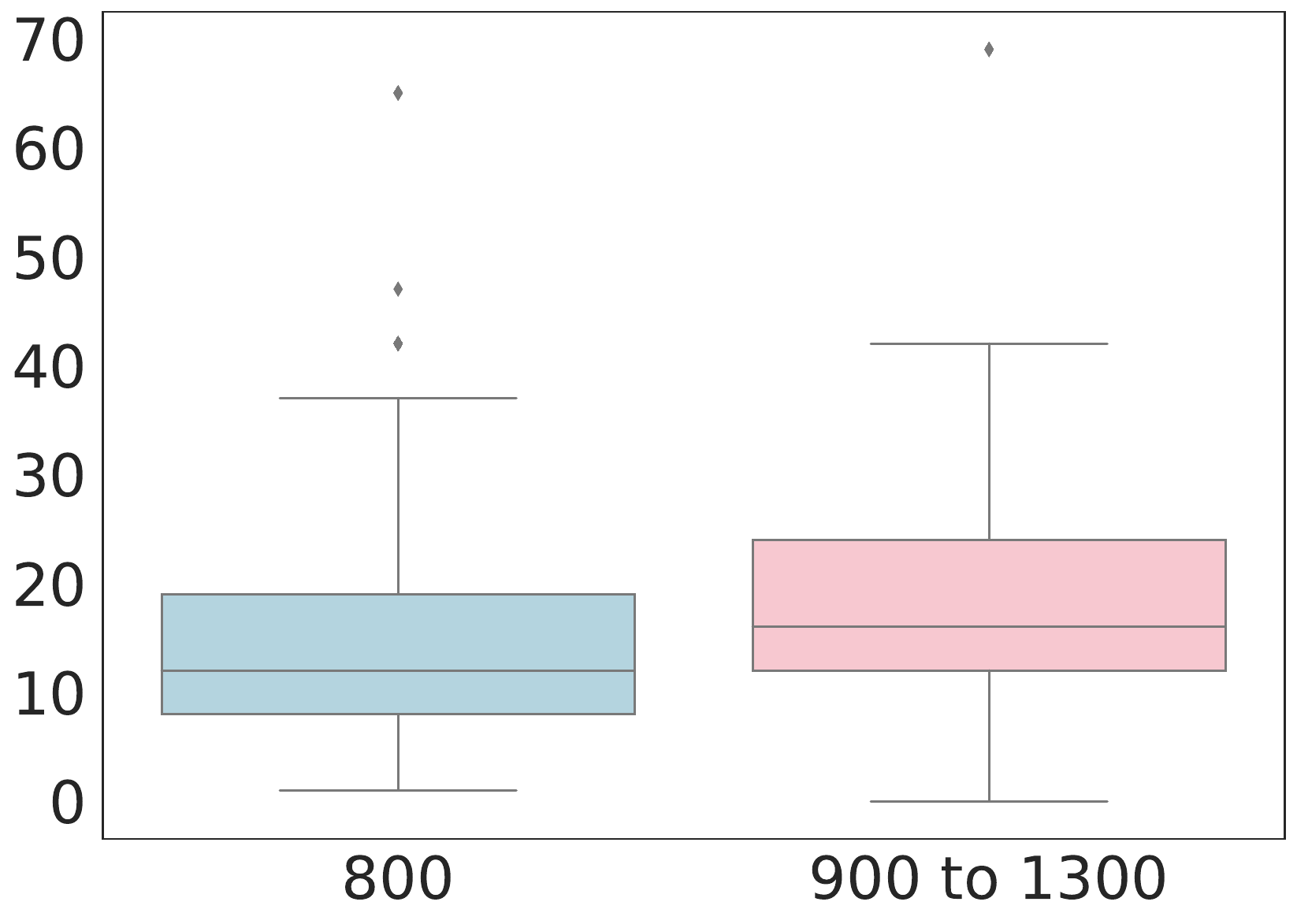}\label{subfig:exprs_difficulty}}
\caption{Number of lines of code and expressions of programs grouped by difficulty}
\label{fig:difficulty}
\end{figure}

The summary statistics of the dataset are shown in Table~\ref{tab:correct_incorrect_num}. The Total column indicates the number of valid comparisons each programming problem comprises, while  the Correct and Incorrect columns display the number of unique programs that are part of the total valid comparisons. The LOC and Exprs.~columns highlight the mean and standard deviation of the number of lines of code and expressions (nodes in the control flow graph) respectively across all programs, both correct and incorrect. The Diff.~column displays the number assigned by Codeforces to indicate the difficulty of the problem. However, the ranking and reasoning behind the assignment of the number are not officially documented. The closest explanation we found was through a Codeforces blogpost~\cite{Mirzayanov_codeforces_nodate}, where difficulty is assigned such that the expected probability of solving the problem is $0.5$ for coders of that rating. Since we are looking at introductory assignments, we limit our ratings from $800$ to $1300$.
We consider the difficulty rating of 800 to be low difficulty (it is the lowest available in Codeforces). We selected the range of 900-1300 as hard difficulty problem as there are 11,841 data points for 800, and 9,518 for 900-1300, allowing for as close to equal binning between the two as possible. Analyzing the table, we observe lower lines of code with greater variance for low-difficulty problems while higher-difficulty problems on average have more lines of code but less variance. This is indicative that low-difficulty problems comprise a combination of one-line and longer solutions, depending on the capability of the programmers. In contrast, the number of expressions at higher difficulty is higher with higher variance, indicating the increase in complexity of the solution to the problems. This is also confirmed by Figure~\ref{fig:difficulty}, which displays boxplots of the number of lines of code and expressions in the programs grouped by difficulty.

\subsection{Quantitative Analysis}

\begin{figure}
\centering
\subfloat[Techniques]{\includegraphics[scale=0.2]{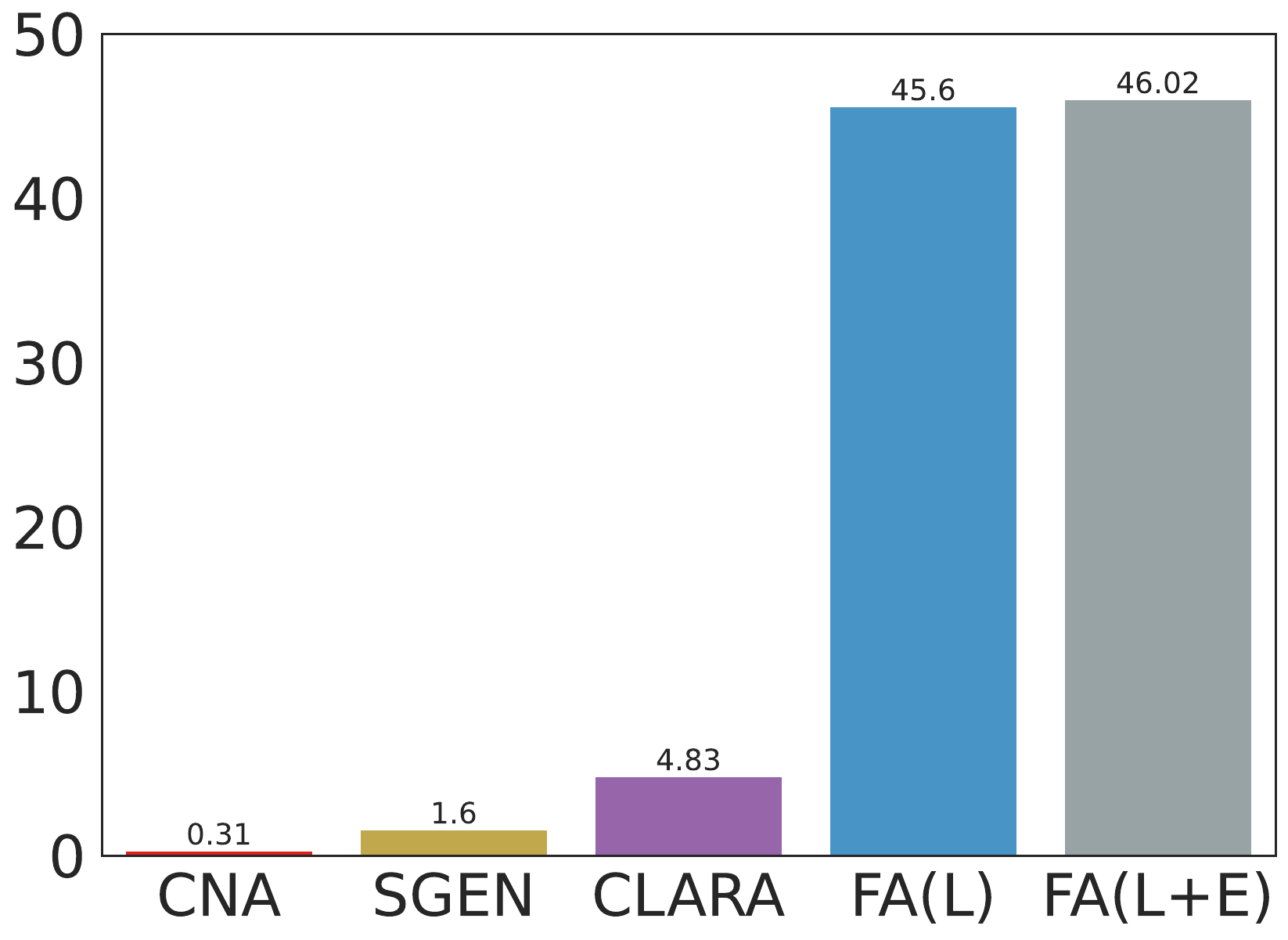}\label{fig:successful_repair_by_experiment}}~
\subfloat[Difficulty]{\includegraphics[scale=0.2]{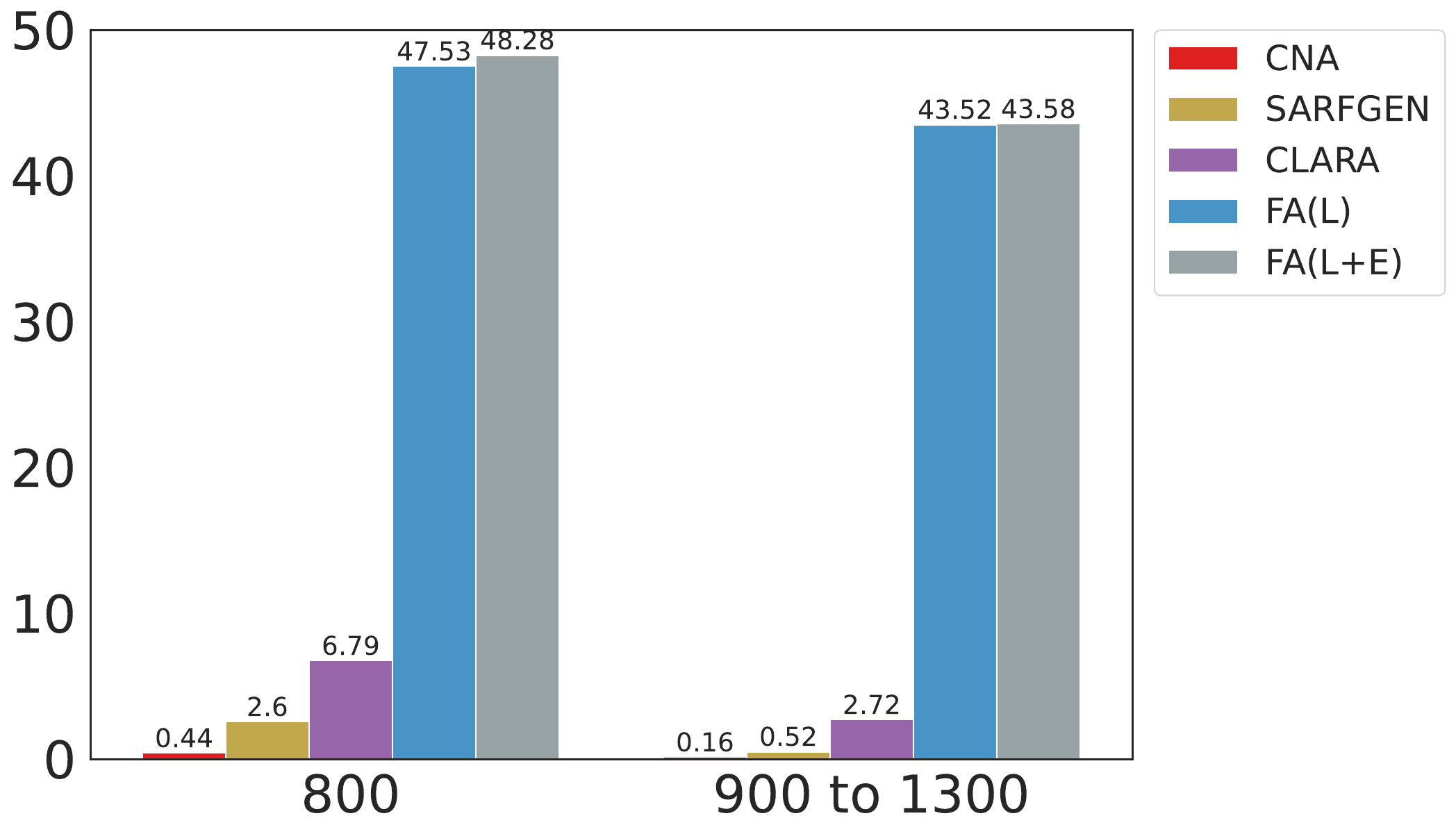}\label{fig:successful_repair_by_diff}}
\caption{Percentage of incorrect programs fully repaired grouped by technique and difficulty}
\label{fig:successful_repair_by_diff_experiment}
\end{figure}

\paragraph{Successful repairs}
In Figure~\ref{fig:successful_repair_by_diff_experiment}, we present the percentage of successful repairs, which was computed as follows: the number of unique incorrect programs fully repaired divided by the total number of incorrect programs. In Figure~\ref{fig:successful_repair_by_experiment}, we present the macro success rate for the \jss{five} techniques under evaluation. It can be observed that flexible alignment in both flavors significantly outperforms CLARA (5\%) and \jss{CNA (0.3\%)} with FA(L) at 45\% and FA(L+E) at 46\% success rates, respectively. \jss{We can also observe that baseline CLARA performs slightly better than SARFGEN and significantly better than CNA.} This macro result highlights the major improvement in repair capability that flexible alignment can achieve. Figure~\ref{fig:successful_repair_by_diff} aggregates results by problem difficulty. We observe that the \jss{five} techniques achieve better success rates in low-difficulty problems. We also observe that success rates decrease in high-difficulty problems compared to low-difficulty ones. In low-difficulty problems, FA(L+E) at 48.2\% outperforms FA(L) at 47.5\% by 0.7\%. However, this gain is not as evident in the other problems. Our hypothesis to explain this behavior is that, on one hand, low-difficulty problems contain similar statements that cannot be easily differentiated based solely on labels. On the other hand, high-difficulty problems contain many specialized statements that are almost unique, so labels are helpful to align statements without the use of edges. We can conclude that using both label and edge similarities in the alignment process is beneficial.


\begin{figure}
\centering
\subfloat[Lines of code]{\includegraphics[scale=0.2]{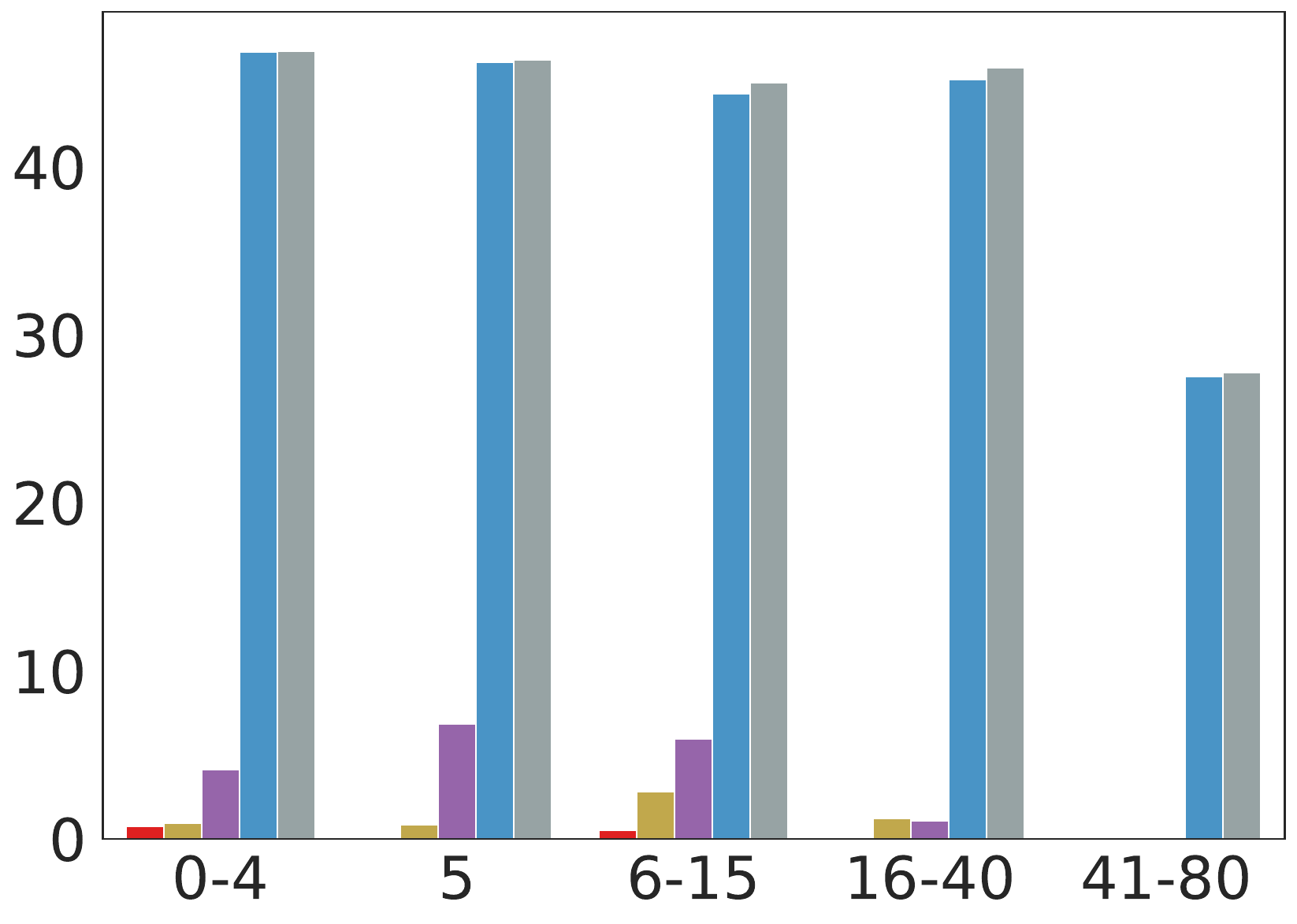}\label{subfig:successful_repair_by_loc}}~
\subfloat[Expressions]{\includegraphics[scale=0.2]{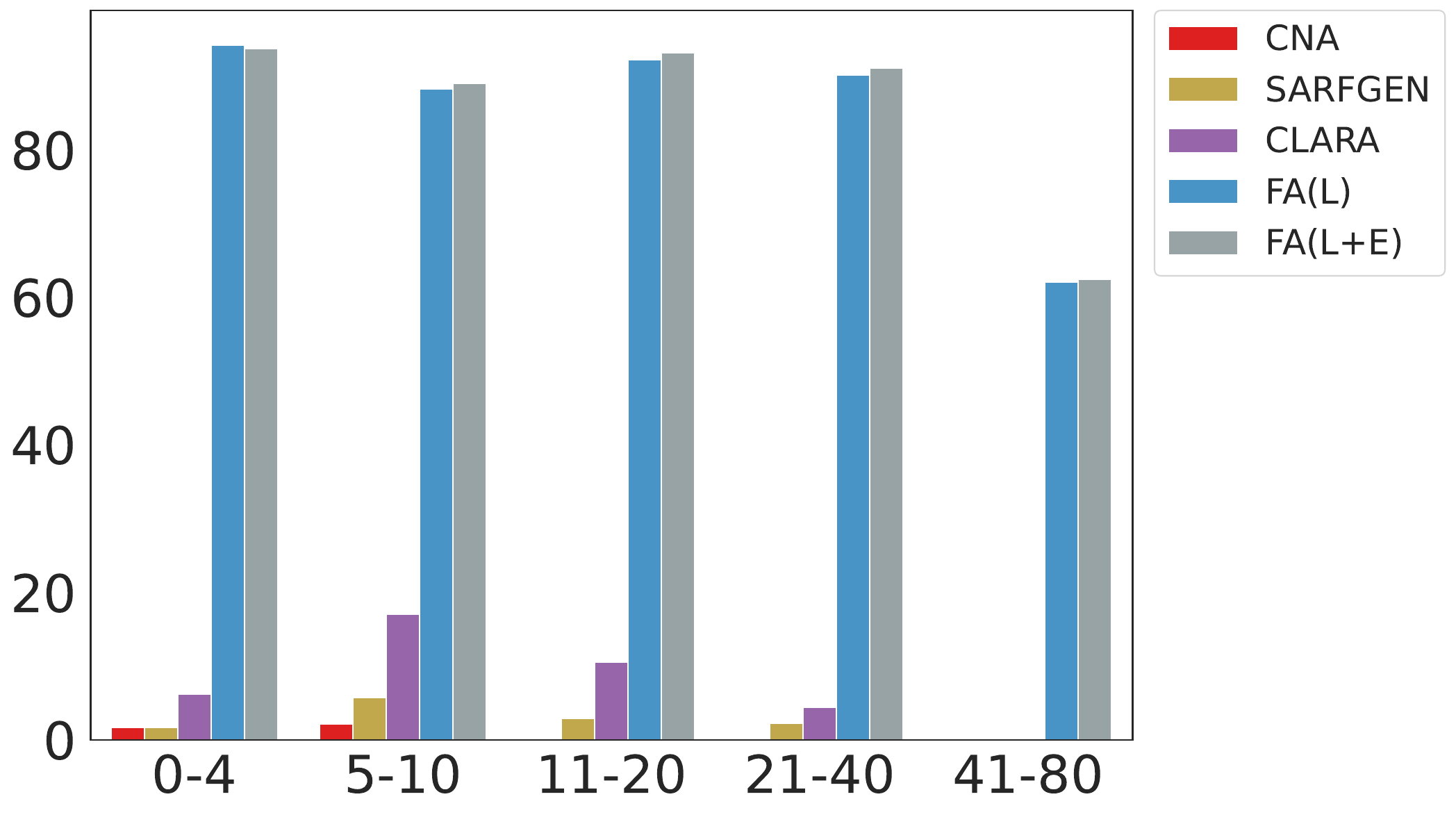}\label{subfig:successful_repair_by_exprs}}
\caption{Percentage of incorrect programs fully repaired grouped by lines of code and expressions. Cases that failed before model creation are ignored.}
\label{fig:successful_repair_by_exprs_loc}
\end{figure}

\begin{table}
\caption{\jss{Number of samples in each bin when grouped by LOC (lines of code) and Exprs.~(expressions). Failed indicates failures before model creation.}}
\centering
\begin{tabular}{|l|c||l|c|}
\hline
LOC & \# & Exprs. & \# \\
\hline
0--4 &      1,072 & 0--4 & 464\\
5 &      2,301 & 5--10 & 1,597\\
6--15 &     2,661 & 11--20 & 3,398 \\
16--40 &    1,712 & 21--40 & 2,230 \\
41--80 &    209 & 41--80 & 266 \\
\hline
Failed & 7,733 & Failed & 7,733 \\
\hline
\end{tabular}
\label{tab:bins}
\end{table}

As lines of code increase for programs as shown in Figure~\ref{subfig:successful_repair_by_loc}, the performance of all the techniques decreases, but baseline CLARA fails to find repairs for programs with 41 lines of code or higher. Note that these bins are not balanced as presented in Table~\ref{tab:bins}. In 7,733 of these comparisons, the techniques failed to produce a model to compute lines of code and expressions, which are reported in the table.
In Figure~\ref{fig:successful_repair_by_exprs_loc}, once model creation has been completed, flexible alignment has a success percentage over 80\% for all LOC and Exprs.~except for those above 41. In Figure~\ref{subfig:successful_repair_by_loc}, FA(L+E) with its additional topological information and flexibility performs better compared to more rigid alignment schemes. In addition to lines of code, if we analyze successful repairs in terms of expressions in programs, FA(L+E) with both semantic and topological information outperforms FA(L), even as the number of expressions -- by proxy complexity -- increases, as seen in Figure~\ref{subfig:successful_repair_by_exprs}.

\begin{figure}
\centering
\subfloat[Number of repairs]{\includegraphics[scale=0.2]{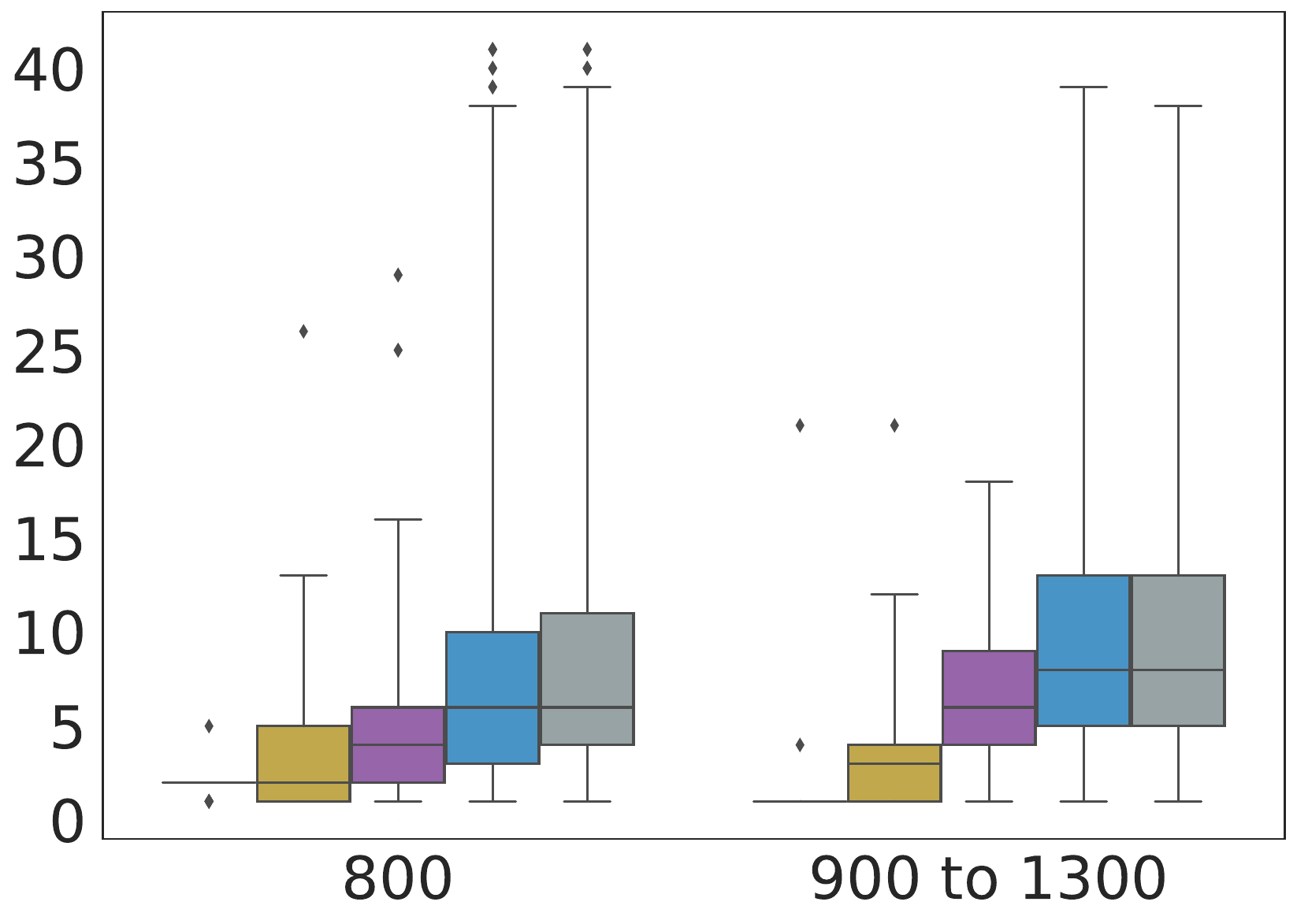}\label{subfig:repairs_by_diff}}
\subfloat[Change percentage]{\includegraphics[scale=0.2]{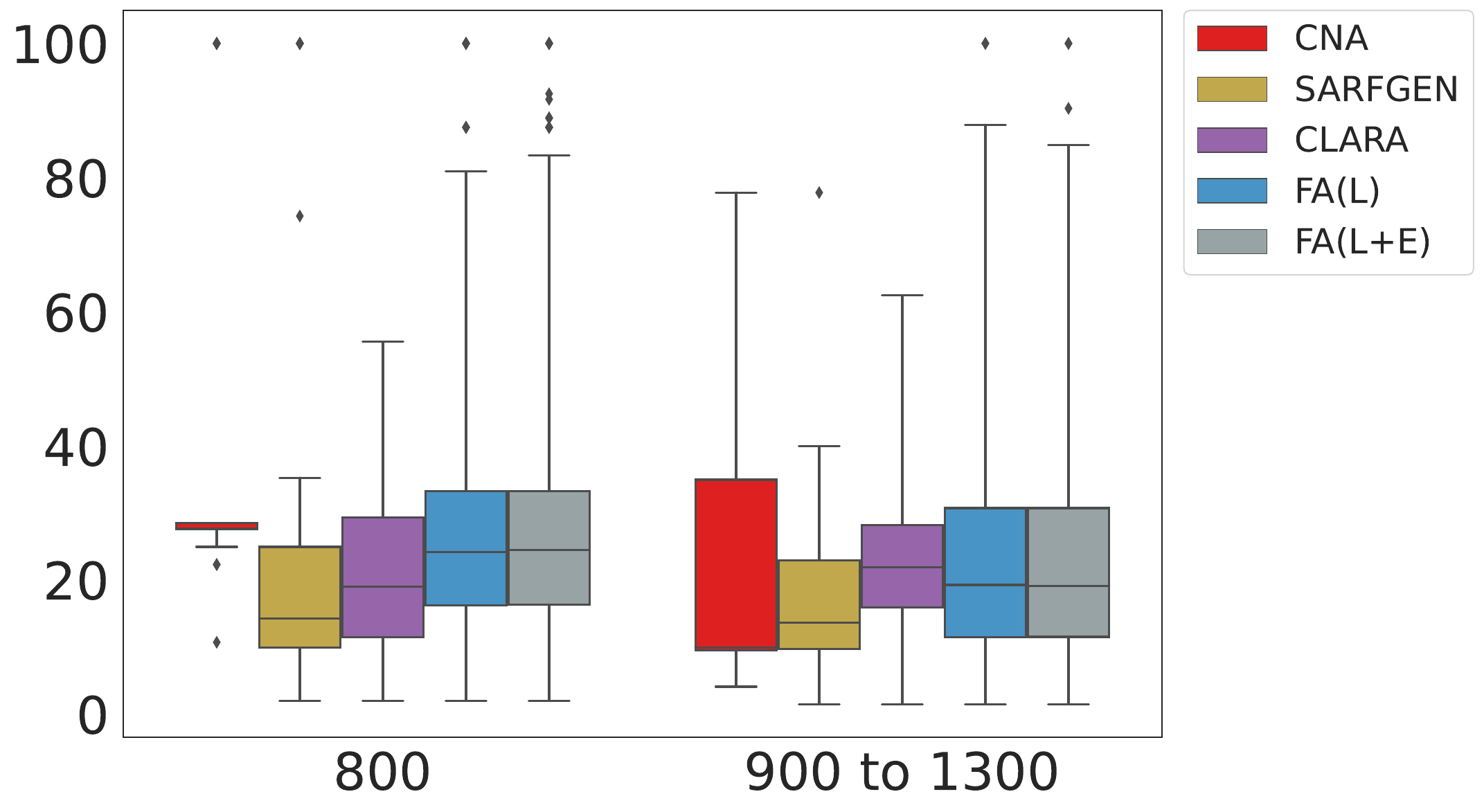}\label{subfig:percentage_repaired_by_diff}}
\caption{Number of repairs necessary to repair incorrect programs and change percentage (the proportion of the incorrect program that was changed) grouped by difficulty}
\label{fig:repairs_percentage_repaired_by_diff}
\end{figure}

\paragraph{Number of repairs and percentage changes}
We also evaluate the performance considering the number of repairs. Note that, for a given incorrect program, there is typically the case that several correct programs can be used to repair it. We measure, for each incorrect program, the minimum number of repairs among all correct programs, and the change percentage, that is, the percentage of the incorrect program that was altered in order to repair it. As shown in Figure~\ref{subfig:repairs_by_diff}, flexible alignment has a higher minimum number of repairs than baseline CLARA, and the range of repairs is significantly higher. This happens in both bins of low- and high-difficulty problems. We observe that the number of repairs in the high-difficulty problems is similar for both FA(L) and FA(L+E); however, for low-difficulty problems, FA(L) has a slightly reduced number of repairs compared to FA(L+E) when they both achieve very similar performance. \jss{CNA and SARFGEN achieve less number of repairs than any of the other techniques, and their performances are quite poor for high-difficulty problems.}

Figure~\ref{subfig:percentage_repaired_by_diff} presents the change percentage, i.e., the percentage of the incorrect program that was replaced with the correct program. As expected, for low-difficulty problems, our flexible alignment techniques change a higher percentage of the incorrect programs than baseline CLARA. Surprisingly, this is not the case in the high-difficulty problems, in which we observe a higher mean of change percentages for baseline CLARA. This implies that our flexible scheme finds smaller repairs than baseline CLARA. The behavior of both FA(L) and FA(L+E) are very similar. This highlights the benefit of using both label and edge flexible alignments for high-difficulty problems. In contrast, SARFGEN significantly reduces the number of changes, but this comes with the penalty of very low repair rates. \jss{For CNA, without any flexible alignment or parser modifications, the median number of changes is lower than SARFGEN but at a penalty of even lower repair rates than SARFGEN due to the rigidity of the approach i.e. it will fix only similar programs.}

\begin{figure}
\centering
\subfloat[Number of repairs]{\includegraphics[scale=0.2]{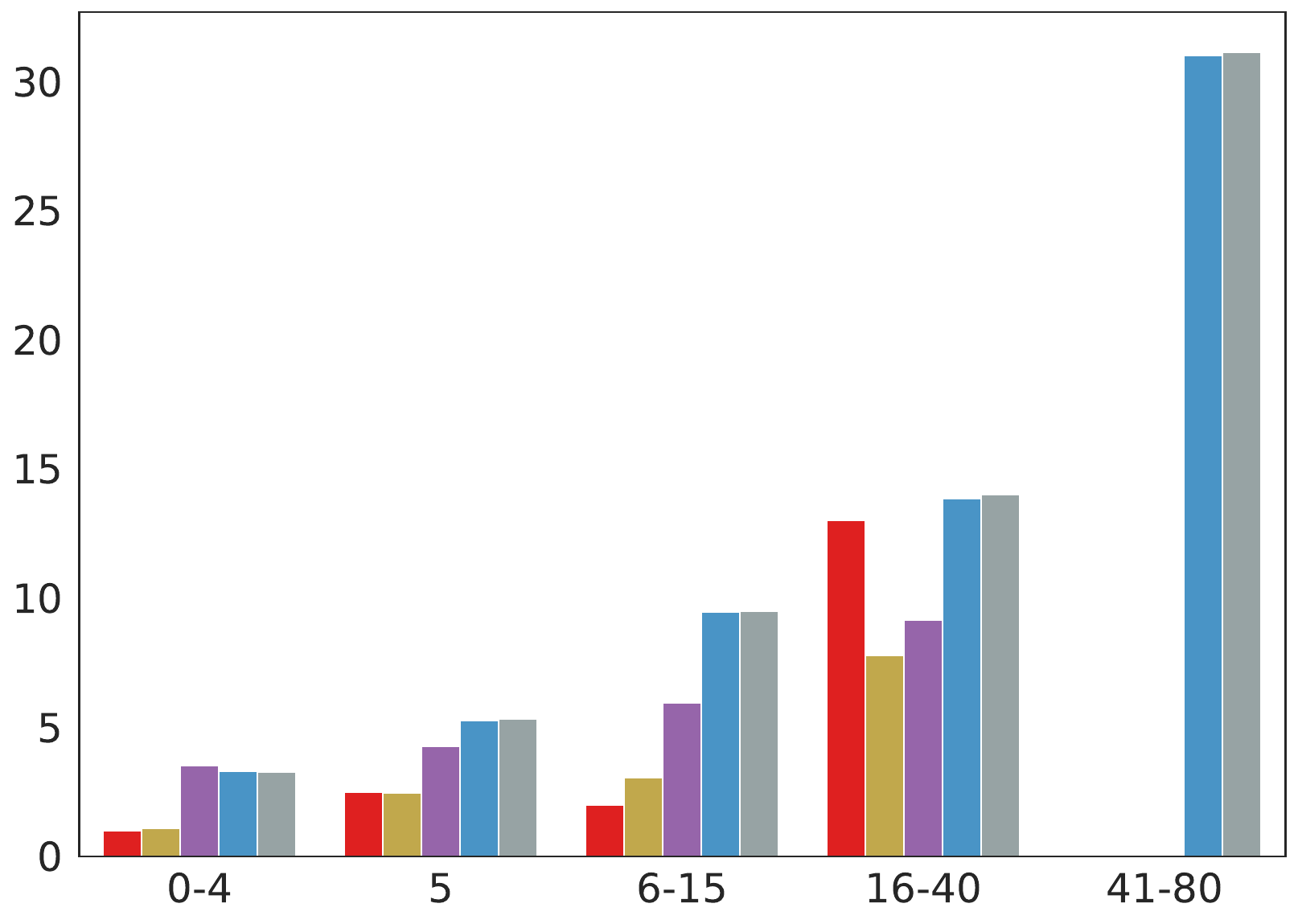}\label{subfig:number_of_repairs_by_loc}}
\subfloat[Change percentage]{\includegraphics[scale=0.2]{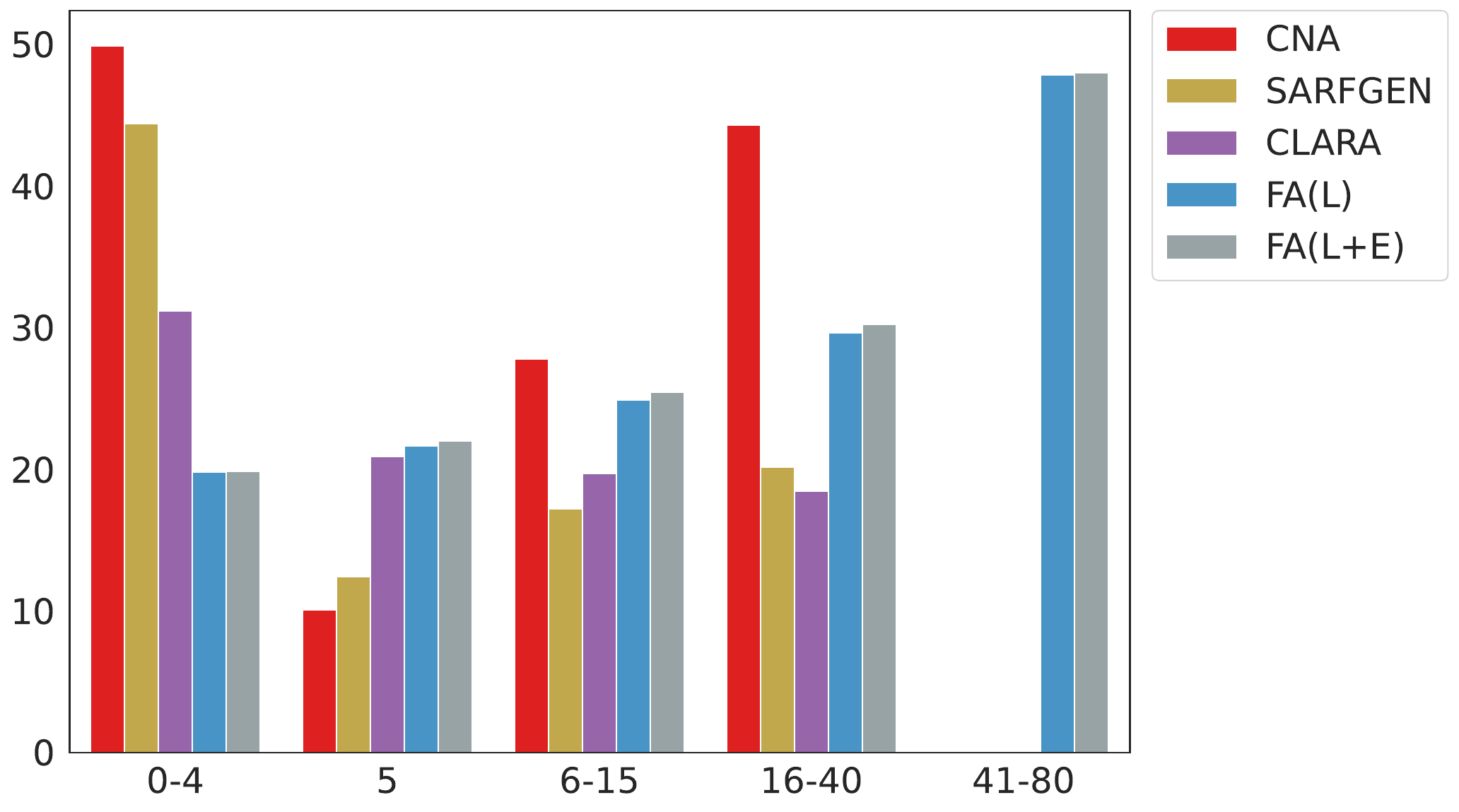}\label{subfig:average_percentage_repaired_by_loc}}
\caption{\jss{Average number of repairs necessary to repair incorrect programs and change percentage (the proportion of the incorrect program that was changed) grouped by lines of code}}
\label{fig:num_repairs_avg_percentage_by_loc}
\end{figure}

Figure~\ref{fig:num_repairs_avg_percentage_by_loc} presents the average number of repairs and change percentage achieved by each technique when grouped by lines of code. We observe that, as lines of code increase, the average change percentage of baseline CLARA decreases, while the same average for our flexible schemes increases. It is surprising though that, for the small bin (0--4), baseline CLARA's and SARFGEN's means are higher than those of our flexible schemes while \jss{CNA has the highest mean}. The presence of labels and edges allows our flexible approach to reduce the change percentage by identifying key labels and edges, while also maintaining a high degree of repairs at higher lines of code. This suggests that by using flexible alignment, one can find a correct--incorrect program comparison that is more efficient than using a rigid program comparison scheme.

\subsection{Qualitative Analysis}
\label{subsec:qual}

\begin{figure}
\centering
\includegraphics[scale=0.23]{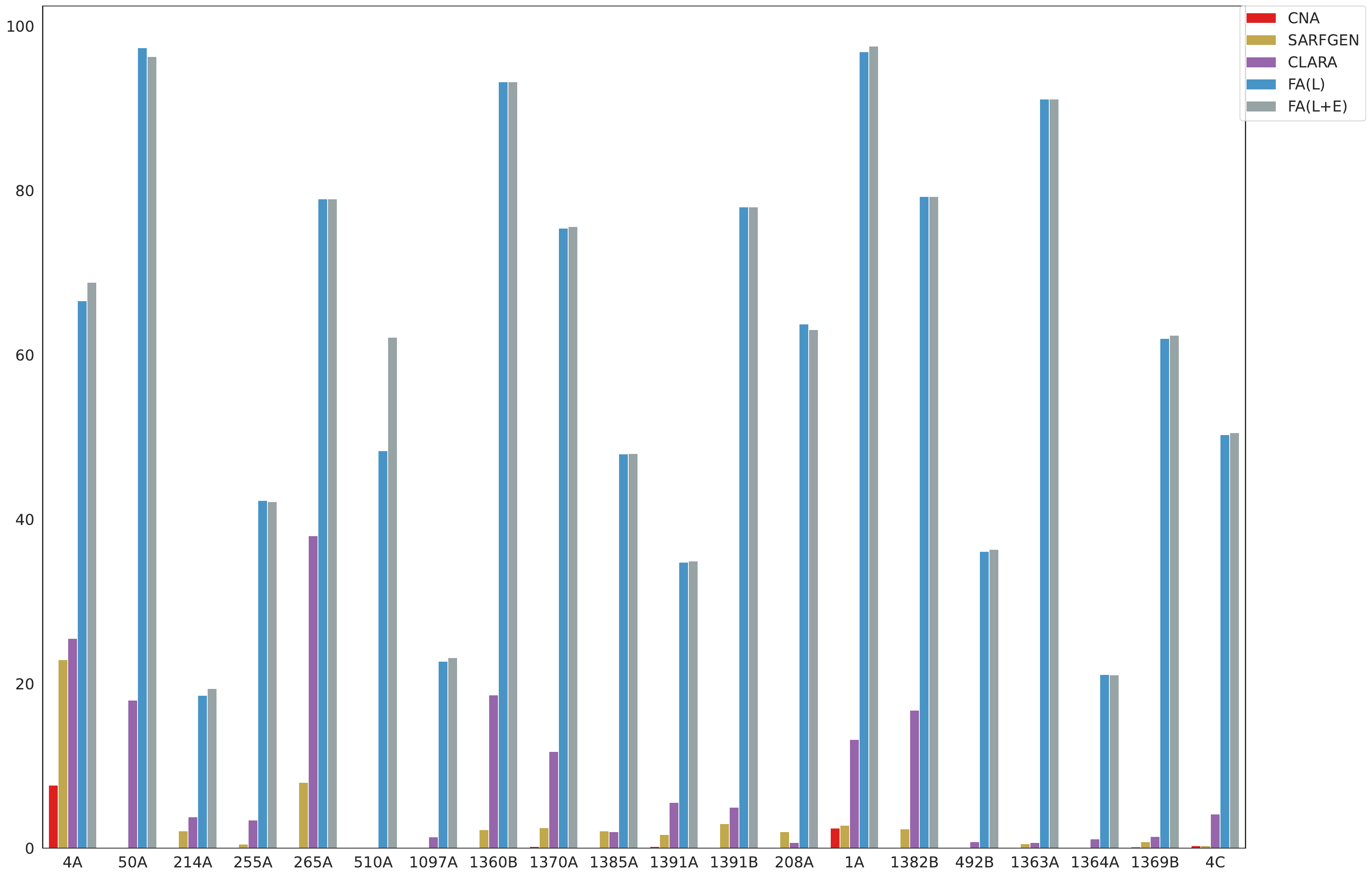}
\caption{Percentage of programs fully repaired grouped by programming problem}
\label{fig:successful_repair_by_problem}
\end{figure}



We conducted a fine-grained analysis of the repairs within the programming problems selected in our dataset. Figure~\ref{fig:successful_repair_by_problem} shows the percentage of programs successfully repaired for each technique grouped by problem. 
The results achieved by \jss{baseline CLARA, CNA, SARFGEN, FA(L), and FA(L+E) across the twenty problems are consistent with the macro results in the previous section.} Flexible alignment consistently outperforms baseline CLARA and CNA: in four problems (50A, 1360B, 1A, 1363A), FA(L) and FA(L+E) achieve more than 80\% of successful repairs, while baseline CLARA achieves less than 20\% success rate. CLARA outperforms SARFGEN except in problems 1385A and 208A. In these two problems, FA(L), and FA(L+E) are far superior. \jss{CNA fails to find solutions in most problems with no flexibility or alignment except for 4A and 1A.}

We observe several problems in which baseline CLARA achieves poor performance of 5\% or less success rate. In the worst cases, FA(L) and FA(L+E) achieve approximately 20\% success rate (problems 214A, 1097A, and 1364A), still outperforming the rest. 
Comparing the performance of FA(L) vs.~FA(L+E), we observe that, in problem 510A, FA(L+E) significantly outperforms FA(L). In the rest of the problems, both techniques perform similarly.

\subsection{Threats to validity}
\label{subsec:threats}

We built our dataset of correct and incorrect programs from Codeforces and utilized their assignment of difficulty as a metric for our comparisons. However, the reasoning and ranking behind the difficulty assignment from Codeforces for each of the problems are not well documented (see discussion above). We present an analysis to check correlations between difficulty, lines of code, and expressions (code complexity) in Figure~\ref{fig:difficulty} to get a better understanding of the dataset. Due to the addition of the graph alignment step, we need to process the feedback returned by CLARA before it is provided to students. Based on the locations added or deleted and the edges modified, we need to be able to inform the learner to add/delete the corresponding expressions.

Another limitation of the flexible alignment step is that, while removing locations from the incorrect program model, it is possible to remove a variable with no other definition in the rest of the program. Therefore, causing the program to crash if it is used. We hope to address this issue in the future by checking if deleting a variable can cause the program to crash and, if so, adding that variable and its expression to a different location.


Recent advances in large language models like ChatGPT\footnote{https://chat.openai.com/} and Codex~\cite{chen2021evaluating} have raised the question of whether they can be used for program repairs such as the one we discuss in this paper. While such language models are good at providing suggestions during coding or to generate introductory code from scratch, they are not yet developed enough to identify and repair incorrect programs given a correct program and test cases as a reference. For example, we prompted ChatGPT with the following request: \textit{Fix the issues in the incorrect code to match the correct code}. The result of ChatGPT was to replace the entire incorrect program with the correct program instead of identifying exactly the issues of the incorrect program. Our future work will focus on adapting our flexible alignment scheme utilizing the semantic strength of such large language models.

\section{Related work}
\label{sec:related}

There are many approaches to automatically repair programs in different areas~\cite{DBLP:journals/csur/Monperrus18}. We categorize these into data-~and non-data-driven. We also discuss program comparison approaches.

\paragraph{Data-driven feedback} CLARA~\cite{DBLP:conf/pldi/GulwaniRZ18} clusters correct programs based on test cases and variable traces. Each incorrect program is compared to the representative of each cluster to find minimal repairs. The repairs consist of adding new variables and modifying existing statements without changing the control flow of the incorrect program. Sarfgen~\cite{DBLP:conf/pldi/WangSS18} searches for correct programs that share the same control flow structure as the incorrect program. Incorrect and correct programs are fragmented based on their control flows, and, for each fragment pair that is matched, potential repairs are computed using abstract syntax tree edits. CLARA and Sarfgen only consider pairs of programs whose control flow match, which is a hard constraint since such a pair may not currently be present in the set of correct programs or, when the incorrect program significantly deviates from a correct program, a correct program with such a control flow may not even be possible.~\citet{claraRefactor} addressed CLARA's drawback of rigid program comparisons by refactoring the correct program at hand using a set of predefined transformations, such that its control flow matches the incorrect program at hand. Using program refactoring, it is possible to modify a program so thoroughly that it no longer resembles the original version, and can potentially cause a correct program to become incorrect. Furthermore, the repairs suggested to change the incorrect program into a correct one need to be backtraced to the original program before refactoring.

Refazer~\cite{DBLP:conf/icse/RolimSDPGGSH17} proposes ``if-then'' rules to match and transform abstract syntax subtrees of a program. Such rules are synthesized from sample pairs of correct/incorrect programs, in which tree edit distance comparisons between correct and incorrect programs help identify individual transformations. Refazer has been extended to propagate feedback based on learned transformations~\cite{DBLP:conf/lats/HeadGSSFDH17}. sk\_p~\cite{DBLP:conf/oopsla/PuNSB16} relies on neural networks to repair incorrect programs. It constructs partial fragments of three consecutive statements using these renamed tokens. The middle statements are removed and fed to the repairer for training. The order of statements is one of the main drawbacks of Refazer, Sarfgen, and sk\_p: Refazer and Sarfgen rely on edit distances of abstract syntax trees, while sk\_p treats programs as documents. Our flexible alignment allows to account for more implementation variability and increases the number of valid program comparisons. \citet{DBLP:conf/icml/PiechHNPSG15} select a subset of existing programs to annotate with feedback. Each annotation is used individually to learn a binary classifier to propagate feedback to unseen programs. These binary classifiers are applied to each incorrect program to decide whether it should be annotated with a piece of feedback. This approach requires the number of variables in programs to be fixed beforehand and a large number of existing programs~\cite{DBLP:conf/icml/PiechHNPSG15}.

\paragraph{Non-data-driven feedback} AutoGrader~\cite{DBLP:conf/pldi/SinghGS13} allows to define rules using an error model language to describe potential repairs to be applied to incorrect programs, e.g., a condition $x < y$ can be mistaken by $x \leq y$. Based on these rules, AutoGrader generates a ``sketch'' of the program, i.e., a program that contains multiple choices for those statements that matched the given rules~\cite{DBLP:journals/sttt/Solar-Lezama13}. A correct program is then assembled by ensuring functional equivalence with respect to a single, reference program. Repairs are computed as the changes to transform from an incorrect to a correct program. There are several approaches that rely on program sketching to compute repairs~\cite{DBLP:conf/cav/DAntoniSS16, DBLP:conf/icse/HuaZWK18, DBLP:conf/icse/Liu0WW19}. Codewebs~\cite{DBLP:conf/www/NguyenPHG14} allows to search for code snippets by exploiting probabilistic semantic equivalence between abstract syntax trees to perform the matching, which is based on functional tests over the trees. Feedback can be propagated to identified code snippets that are similar.~\citet{DBLP:conf/icde/MarinPSR17} encode correct and incorrect feedback in subgraph patterns over program dependence graphs. Feedback is propagated based on exact subgraph matching with approximations at the statement level defined by regular expressions. Verifix~\cite{DBLP:journals/tosem/AhmedFYAR22} uses satisfiability modulo theories solvers to find verified repairs between incorrect and correct programs.~\citet{DBLP:conf/icse/EdmisonE20},~\citet{DBLP:conf/icsm/NguyenLLDL0H22} and~\citet{DBLP:journals/jss/LiWLSWZC23} applied fault localization techniques to detect defects in student programs and suggest repairs.

Many approaches have focused on discovering repairs by mutating programs until repairing them~\cite{DBLP:journals/csur/Monperrus18, DBLP:conf/kbse/HuAMLR19}. These mutations can be predefined and explored using genetic algorithms~\cite{DBLP:journals/tse/GouesNFW12}.~\citet{DBLP:conf/sigsoft/YiAKTR17} analyzed the usage of some of these approaches to repair student programs, and concluded that they are better suited for programs that fail a small number of tests, while student programs are typically significantly incorrect. Mutations can also be retrieved from existing software repositories~\cite{DBLP:conf/popl/LongR16, DBLP:conf/pldi/Sidiroglou-Douskos15, DBLP:conf/kbse/XinR17}. While these approaches can be seen as data-driven, they aim to find repairs based on programs that are generally not related to the incorrect program at hand to be repaired; therefore, this is a more difficult problem than the one we aim to tackle.

\paragraph{Program comparison} There is a large body of knowledge of program comparison in the context of code clones and code plagiarism, i.e., copied-and-pasted pieces of code with some possible modifications. Successful code clone detectors have focused on comparing program tokens and (features of) abstract syntax trees~\cite{DBLP:journals/scp/RoyCK09}. These detectors find blocks of lines of code that are similar, but they usually fail to detect correspondences between statements. Program dependence graphs are believed to achieve the best accuracy when detecting Type 4 clones, i.e., two pieces of code that perform the same computation but are implemented by different syntactic variants~\cite{DBLP:journals/scp/RoyCK09}. Existing approaches have mainly focused on comparing programs based on subgraph isomorphism~\cite{DBLP:conf/kdd/LiuCHY06, DBLP:conf/icse/LiE12, DBLP:conf/issre/SunSPLZY10, DBLP:journals/tse/XuC03}; however, they are generally not flexible enough to cope with implementation variability, and they only provide binary comparisons (Are graphs isomorphic? Is a graph contained in the other graph?).

The approach by~\citet{DBLP:conf/icsr/LiSSSS16}, the most related to our approximate alignment, compares the kernel representations of data-flow and API-call graphs. In this case, a kernel is the histogram of node colors that result after the Weisfeiler-Leman algorithm is applied for several rounds. This algorithm computes an initial coloring for each node based on its immediate neighbors, which is later refined in subsequent rounds. It only computes graph topological similarity while our approach aims to combine both topological and semantic similarities of nodes. Also, this approach does not compute similarities between mapped nodes that can be later exploited.

\section{Conclusions}
\label{sec:conclusions}

Nowadays, programming is perceived as a must-have skill. It is thus not surprising that the number of learners have scaled to millions, especially in online settings. Delivering feedback is addressed by repairing learners' incorrect programs. The trend in data-driven approaches is to perform a rigid matching between correct and incorrect programs to discover snippets of code with mending capabilities. The downside is that potential repairs that could be captured by looser alignments may be missed.

This paper explores using a flexible alignment between statements in pairs of programs to discover potential repairs. We extend an existing data-driven automated repair approach that is open source, CLARA, with our flexible alignment approach to deal with such real-world problems. We utilize the abstract syntax tree parser in Python to build control flow graphs, and assign a similarity to aligning a node in a correct program to a node in an incorrect program. 
In our evaluation, we compare flexible alignment with respect to rigid program comparisons. The former is capable of repairing more programs than rigid schemes, which supports our hypothesis that rigid approaches might be missing valuable code snippets for repairs that could be discovered by an approximate method otherwise. Furthermore, our analysis reveals that flexible alignment also decreases the changes required to fix more difficult problems. For shorter programs, less number of changes are necessary than when using rigid schemes. 
As a result, we claim that ``search, align, and repair'' approaches should rely on flexible alignments to improve their repair capabilities. 
Our analysis comparing both semantic and topological (labels and edges) similarities of our flexible alignment approach indicates that using both types of similarities is beneficial compared to only using labels. 

In future work, we plan to integrate our flexible alignment schemes with repairs based on variable traces or program sketches. We will use other node semantic similarities rather than the Jaccard distance between multisets of labels, such as graph representation of statements. We also plan to study how large language models can be leveraged to improve our flexible alignment approach. 

\bibliographystyle{elsarticle-num-names} 
\bibliography{cas-refs}





\end{document}